# SN 2019nyk: A rapidly declining Type II supernova with early interaction signatures

Raya Dastidar[1,2,⋆], Giuliano Pignata[3], Naveen Dukiya[4,5], Kuntal Misra[4], Daichi Hiramatsu[6,7,8,9], Javier Silva-Farfán[10], D. Andrew Howell[8,9], K. Azalee Bostroem[11], Mridweeka Singh[12], Anjasha Gangopadhyay[13], Amit Kumar[14], Curtis McCully[8,9]

[1] Instituto de Astrofísica, Universidad Andres Bello, Fernandez Concha 700, Las Condes, Santiago RM, Chile
[2] Millennium Institute of Astrophysics (MAS), Nuncio Monsenor Sòtero Sanz 100, Providencia, Santiago RM, Chile
[3] Instituto de Alta Investigación, Universidad de Tarapacá, Santiago RM, Chile
[4] Aryabhatta Research Institute of observational sciencES, Manora Peak, Nainital, 263001, India
[5] Department of Applied Physics, Mahatma Jyotiba Phule Rohilkhand University, Bareilly, 243006, India
[6] Center for Astrophysics | Harvard & Smithsonian, 60 Garden Street, Cambridge, MA 02138-1516, USA
[7] The NSF AI Institute for Artificial Intelligence and Fundamental Interactions, USA
[8] Las Cumbres Observatory, 6740 Cortona Drive Suite 102, Goleta, CA, 93117-5575, USA
[9] Department of Physics, University of California, Santa Barbara, CA 93106-9530, USA
[10] Departamento de Astronomía, Universidad de Chile, Casilla 36D, Santiago, Chile
[11] DiRAC Institute, Department of Astronomy, University of Washington, Box 351580, U.W., Seattle, WA 98195, USA
[12] Indian Institute of Astrophysics, Koramangala 2nd Block, Bangalore 560034, India
[13] Hiroshima Astrophysical Science Center, Hiroshima University, Higashi-Hiroshima, Japan
[14] Department of Physics, University of Warwick, Coventry CV4 7AL, UK

Received ; accepted

**ABSTRACT**

We present an optical photometric and spectroscopic analysis of the fast-declining hydrogen-rich Type II supernova (SN) 2019nyk. The light curve properties of SN 2019nyk align well with those of other fast-declining Type II SNe, such as SNe 2013by and 2014G. SN 2019nyk exhibits a peak absolute magnitude of $-18.09 \pm 0.17$ mag in the $V$ band, followed by a rapid decline at $2.84 \pm 0.03$ mag $(100\,\text{d})^{-1}$ during the recombination phase. The early spectra of SN 2019nyk exhibit high-ionisation emission features as well as narrow H Balmer lines, persisting until 4.1 d since explosion, indicating the presence of circumstellar material (CSM) in close proximity. A comparison of these features with other Type II SNe displaying an early interaction reveals similarities between these features and those observed in SNe 2014G and 2023ixf. We also compared the early spectra to literature models, estimating a mass-loss rate of the order of $10^{-3}$ $\text{M}_\odot$ $\text{yr}^{-1}$. Radiation hydrodynamical modelling of the light curve also suggests the mass loss from the progenitor within a short period prior to explosion, totalling $0.16\,\text{M}_\odot$ of material within $2900\,\text{R}_\odot$ of the progenitor. Furthermore, light curve modelling infers a zero-age main sequence mass of $15\,\text{M}_\odot$ for the progenitor, a progenitor radius of $1031\,\text{R}_\odot$, and an explosion energy of $1.1 \times 10^{51}$ erg.

**Key words.** supernovae: general – supernovae: individual: SN 2019nyk – supernovae: individual: ZTF19abqrhvt

## 1. Introduction

Massive stars ($\gtrsim 8\,\text{M}_\odot$) typically lose varying amounts of their outer envelope mass ($\dot{M} > 10^{-6}$ $\text{M}_\odot$ $\text{yr}^{-1}$) through radiation-driven winds, eruptive mass-loss events, or a binary interaction during their evolution from the Zero Age Main Sequence (ZAMS) to the onset of core collapse (see Smith 2014 for a review and references therein). Hydrogen-rich type II supernovae (SNe) are known to originate from massive stars that have not undergone significant mass loss from the outer hydrogen (H) layer prior to the collapse, as evidenced by strong H Balmer signatures in their optical spectra. With an early detection and rapid response spectroscopic observation, we can place constraints on the mass lost by the progenitor shortly before the explosion (e.g. Gal-Yam et al. 2014). Recent advancements in observational capabilities have facilitated the early detection of numerous transient events. This progress has been driven by high-cadence all-sky surveys such as the Zwicky Transient Facility (ZTF, Bellm et al. 2019), Panoramic Survey Telescope and Rapid Response System (Pan-STARRS, Chambers et al. 2016), Asteroid Terrestrial-impact Last Alert System (ATLAS, Tonry et al. 2018), and Gravitational-wave Optical Transient Observer (GOTO, Steeghs et al. 2022). These surveys are complemented by the rapid dissemination of alerts through community alert brokers such as Automatic Learning for the Rapid Classification of Events (ALeRCE) (Förster et al. 2021).

The class of H-rich Type II SNe is diverse, featuring light curves (LCs) that exhibit a continuous range of post-maximum decline rates. These rates encompass both slow-declining events, traditionally referred to as IIP ('P'lateau), and fast-declining events, also known as IIL ('L'inear) (Barbon et al. 1979; Anderson et al. 2014; Valenti et al. 2016). LCs of these SNe typically consist of an initial optically thick phase driven by H recombination, followed by a subsequent optically thin phase powered by the radioactive decay of $^{56}$Co. The increasing decline rate of the LCs is suggested to arise from progenitors with a progressively lower retained H-envelope mass prior to explosion (Georgy 2012). Various criteria have been proposed in the existing literature to quantitatively differentiate between slow and fast-declining Type II SNe. However, in this paper, we refrain from making such distinctions. There are dozens of SNe II on the slower declining end of the continuum for which analysis of high resolution pre-explosion images could constrain a red supergiant (RSG) progenitor in the mass range of ≈ 8 - 18 $M_\odot$ (Smartt et al. 2009; Maund et al. 2014; Smartt 2015; Van Dyk 2017 and references therein, however, see also Davies & Beasor 2018 suggesting the mass range might extend up to 19 $M_\odot$ or even higher). On the other hand, studies based on pre-explosion images of Type II SNe 2009hd and 2009kr, which correspond to the faster declining end of the continuum, could only indicate progenitor mass upper limits: <20 $M_\odot$ for SN 2009hd (Elias-Rosa et al. 2011) and <25 $M_\odot$ for SN 2009kr (Maund et al. 2015).

⋆ E-mail: rdastidr@gmail.com

The primary criterion for classifying SNe as Type II is the presence of prominent H Balmer P Cygni profiles in their maximum light spectrum. However, in the pre-maximum spectra of several Type II SNe acquired shortly after the explosion, narrow emission features can be observed superimposed on a blue continuum. These features are the high ionisation emission lines of helium, carbon, nitrogen, and oxygen, commonly referred to as 'flash features' in the literature. The widths of these lines correspond to a velocity dispersion of ~100 - 1000 km s$^{-1}$ (e.g. Groh 2014; Yaron et al. 2017). Their existence is transitory, lasting only briefly following the explosion. These features form in the unshocked circumstellar material (CSM) and are therefore only observable in early spectra before the SN ejecta overrun the dense CSM.

The dense CSM close to the progenitor likely formed during a super-wind phase (Quataert & Shiode 2012; Fuller 2017) occurring shortly before the core collapse (Khazov et al. 2016; Yaron et al. 2017). However, alternate possibilities for the formation of a dense CSM have been proposed, such as progressive mass overloading taking place over the entire RSG lifetime (Dessart et al. 2017; Soker 2021), RSG envelope convection and associated instabilities (Goldberg et al. 2022; Kozyreva et al. 2022), or colliding-wind binaries (Kochanek 2019). In the study of Khazov et al. (2016), up to 18 per cent of young SNe, with spectra obtained within 5 days of explosion, have exhibited high ionisation emission lines. Additionally, Bruch et al. (2023) found that 36 per cent of young SNe observed within 2 days of explosion displayed such features.

These early spectra offer valuable insights into various aspects of the SN progenitor, including its mass-loss rate, wind velocities, and the chemical composition of its surface layer. There are Type II SNe for which the narrow emission features in the spectra persist for extended periods, typically months to years, and these are designated as SNe IIn. While the presence of CSM has a considerable impact on Type IIn SNe throughout their evolution, the Type II SNe with transient IIn-like features are only briefly impacted by the CSM, from a few hours (e.g. SN 2013fs, Yaron et al. 2017) to a few days (e.g. SN 2023ixf, Bostroem et al. 2023; Smith et al. 2023), transforming into regular Type II SNe within a week. In fact, Dessart et al. (2017) referred to the latter as the circumstellar envelope adjoining the progenitor's surface (within 10 $R_\star$, where $R_\star$ denotes the progenitor radius), rather than a standard CSM (detached material lying between 10 - 100 $R_\star$). Furthermore, we note that the formation process of narrow, high ionisation emission features in early Type II SN spectra mirrors that of interacting Type IIn SNe. Besides the ionising radiation during shock breakout, a sustained ionising flux is necessary to maintain the high ionisation features over several days. The continuous photo-ionisation, induced by the forward shock passing through the pre-shock CSM, can maintain these features for an extended duration. Consequently, these features are referred to as 'IIn-like' in Dessart & Jacobson-Galán (2023) rather than 'flash features'.

In addition to the early spectroscopic features, the influence of ejecta-CSM interaction becomes evident in the luminosity evolution of the SN (e.g. Morozova et al. 2017; Förster et al. 2018; Khatami & Kasen 2023). Khatami & Kasen (2023) have suggested that, in certain interaction scenarios, IIn-like signatures might not emerge in the early spectra, which depends on factors such as the shock-breakout location and CSM density. Nevertheless, the presence of a dense CSM could still manifest in the luminosity evolution, such as a steep rise to maximum in the LC (Förster et al. 2018), which is a signature of wind shock breakout, or an excess luminosity at early times resulting from the conversion of the ejecta's kinetic energy into radiation.

The set of published H-rich Type II SNe exhibiting early interaction signatures in the spectra is quite limited, encompassing only a handful of cases: SNe 2013fs (Yaron et al. 2017; Bullivant et al. 2018), 2013fr (Bullivant et al. 2018), 2014G (Terreran et al. 2016), 2015bf (Lin et al. 2021), 2020pni (Terreran et al. 2022), 2020tlf (Jacobson-Galán et al. 2022), and 2022jox (Andrews et al. 2023). The latest addition to this set is the nearby SN 2023ixf (Jacobson-Galán et al. 2023; Hosseinzadeh et al. 2023; Jencson et al. 2023; Bostroem et al. 2023; Teja et al. 2023; Hiramatsu et al. 2023, and references therein) in the galaxy M101. This work presents the photometric and spectroscopic analysis of another Type II SN, SN 2019nyk, which was observed up to the radioactive tail phase and has manifested IIn-like features in its early spectra.

The structure of the paper is organised as follows: Section 2 focuses on constraining the explosion epoch and certain host galaxy properties, including distance and reddening. Data reduction procedures are outlined in Section 3. In Section 4, we delve into the analysis of the LC, colour curve and the estimation of the $^{56}$Ni mass ejected in the explosion. The subsequent section, Section 5, is dedicated to the spectral analysis, encompassing a comparison of the spectral characteristics with those of other SNe. Explosion parameters, including explosion energy and ejecta mass, are determined through bolometric LC modelling in Section 6. Lastly, the paper concludes with a discussion in Section 7.

## 2. Supernova and host galaxy parameters

### 2.1. Discovery and explosion epoch

SN 2019nyk (a.k.a ZTF19abqrhvt, ATLAS19syc, PS19eup) was discovered by ZTF with coordinates at J2000 RA = 00:15:15.205 and DEC = −08:11:21.84 on 2019-08-20.4 (JD 2458715.92) in the host galaxy HIPASS J0015-08. It was reported by the ALeRCE broker team (Forster 2019). The spatial location of the SN within the host galaxy is depicted in Fig. 1. ALeRCE's stamp classifier (Carrasco-Davis et al. 2021) classified the event as an early SN candidate. The discovery magnitude of the SN was 17.48 ABMag in the ZTF *r*-filter, observed using the Palomar 1.2m Oschin telescope. The last non-detection was reported on 2019-08-15.4 (JD 2458710.92) at a limiting magnitude of 19.48 ABMag in the ZTF *r*-filter. Additionally, we downloaded the forced photometry data from ZTF Forced Photometry Service (ZFPS, Masci et al. 2023) and independently derived the magnitudes for SN 2019nyk. From this, we found the last non-detection was one day later, that is, on 2019-08-16.4 (JD 2458711.90), with a limiting magnitude of 19.29 AB-Mag in the ZTF *r*-filter. A spectrum of the SN was obtained by the extended Public ESO Spectroscopic Survey of Transient Objects team (ePESSTO+, Smartt et al. 2015) on 2019-08-21.2 (JD 2458716.7) using EFOSC2 and Grism 13 (3985-9315 Å, 18 Å resolution) mounted on the ESO New Technology Telescope (NTT) at La Silla. This spectrum was reported to exhibit faint Balmer and He II emission lines superimposed on a blue continuum, leading to its classification as SN II (Fraser et al. 2019). We adopt the mid-point between the last non-detection determined from forced photometry and the first detection as the explosion epoch of SN 2019nyk, which is JD 2458713.9±2.0.

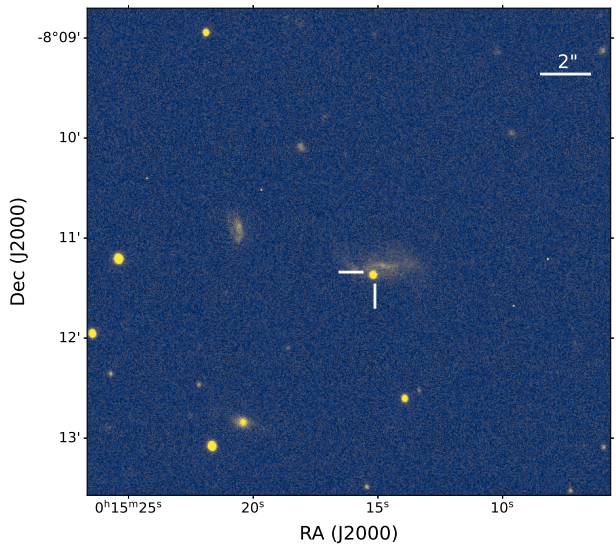

**Fig. 1.** 90s *V* band image of the SN field obtained with 1m LCO telescope is shown and the location of SN 2019nyk in the host galaxy HIPASS J0015-08 is marked.

### 2.2. Host properties

The host galaxy of SN 2019nyk is identified to be HIPASS J0015-08, which was discovered in the HI Parkes All-Sky Survey (HIPASS, Barnes et al. 2001). The published redshift of the host galaxy in Parkash et al. (2018) is 0.021. We use the luminosity distance to the host galaxy, 94.7 ± 7.4 Mpc ($\mu$ = 34.88±0.17 mag), which is the Virgo infall corrected cosmological distance using z = 0.021, provided in the HyperLEDA database and scaled to the value of $H_0$ = 67.4±0.5 km s$^{-1}$Mpc$^{-1}$ (Planck Collaboration et al. 2020).

The Galactic extinction is estimated using the dust extinction model from O'Donnell (1994) with $R_v$ = 3.1 and the Galactic reddening value E(B − V)$_{MW}$ = 0.0327±0.0010 mag, obtained from the extinction dust maps of Schlafly & Finkbeiner (2011). There is no detectable narrow Na ID absorption in the spectra at the redshift of the host galaxy. Hence, we do not include any contribution to reddening of the SN from the host.

We compare some of the properties of HIPASS J0015-08, such as star formation rate (SFR) and stellar mass, with a sample of SN II hosts from the Palomer Transient Factory (Schulze et al. 2021) and metallicity with the PMAS/PPak Integral-field Supernova hosts COmpilation (PISCO, Galbany et al. 2018), to check its consistency with other SN II hosts.

1. The absolute *B* band magnitude of HIPASS J0015-08, is −19.72±0.53 mag (from HyperLEDA), which is similar to the modal value for SN II hosts, −19.75$^{+0.34}_{-0.24}$ mag (Schulze et al. 2021).

2. The stellar mass and SFR of HIPASS J0015-08 are estimated in Parkash et al. (2018) to be $\log_{10}$ M$_*$ = 9.26±0.20 (M$_\odot$) and $\log_{10}$ SFR = −0.86±0.07 (M$_\odot$ yr$^{-1}$), which yields a specific star formation rate (sSFR), $\log_{10}$ sSFR = −10.12±0.21 (yr$^{-1}$). The median values of these parameters in the sample of host galaxies of SN II in Schulze et al. (2021) are $\log_{10}$ M$_*$ = 9.65 ± 0.05 (M$_\odot$), $\log_{10}$ SFR = −0.24±0.04 (M$_\odot$ yr$^{-1}$) and $\log_{10}$ sSFR = −9.86±0.02 (yr$^{-1}$). Thus the sSFR of HIPASS J0015-08 is marginally lower than the sample of SN II hosts.

3. Since both M$_*$ and SFR are correlated with metallicity, we employ equation 2 from Yates et al. (2012) to estimate the host metallicity, resulting in a value of 8.63 ± 0.07 dex (± 0.04 dex, intrinsic scatter). This estimate is comparable to solar metallicity (8.66 ± 0.05 dex, Asplund et al. 2006), however, higher than the median metallicity of 8.54±0.04 dex observed for SN II hosts (Galbany et al. 2018). Nevertheless, the latter study was conducted on a targeted sample, and thus the median metallicity of SN hosts could be higher than the value reported in Galbany et al. (2018).

## 3. Data reduction

Observations of SN 2019nyk in the optical *UBgVri* bands were initiated approximately 2 days after its discovery. These observations were carried out using the Sinistro cameras mounted on the 1-metre telescopes of the Las Cumbres Observatory (LCO) through the Global Supernova Project (GSP) collaboration. The observations spanned a duration of up to 139 days since discovery. Pre-processing of the images, including bias correction and flat-fielding, was conducted using the BANZAI pipeline (McCully et al. 2018). Subsequent data reduction utilised the `lcogtsnpipe` pipeline (Valenti et al. 2016), which is a photometric reduction pipeline built upon PyRAF. This pipeline calculates zero-points and colour terms and subsequently extracts magnitudes using the point-spread function technique (Stetson 1987). The photometry in *UBV* bands is presented in Vega magnitudes, while the *gri* bands data are presented in AB magnitudes, calibrated using SDSS sources (Smith et al. 2002). Given the proximity of SN 2019nyk to its host galaxy, host galaxy contamination was removed by subtracting template images acquired after the supernova had faded, using the PyZOGY image subtraction algorithm (Zackay et al. 2016; Guevel & Hosseinzadeh 2017). The final calibrated photometric data from LCO are compiled in Table 1. We have also included the ZTF *g* and *r* bands forced photometry data in this study. ZTF photometry is on the AB magnitude scale, calibrated using Pan-STARRS1 Survey (PS1) sources (Masci et al. 2019). To bring the LCO and ZTF *g* and *r* bands data into the same system, we used the colour coefficients from Tonry et al. (2012) to convert the ZTF magnitudes from the PS1 to SDSS photometric system. To account for the systematic differences between LCO and ZTF photometry (in SDSS photometric system), an average shift was applied to the ZTF photometry before merging them. We interpolated the LCO *g* and *r* band LCs to the epochs of ZTF photometry and estimated an average shift (difference between the interpolated data and ZTF data) of <$\delta$g> = -0.01 ± 0.07 and <$\delta$r> = 0.07 ± 0.08 mag.

Spectroscopic observations of SN 2019nyk were initiated ∼ 2 days after the discovery with the FLOYDS spectrograph on the 2-m Faulkes Telescope North and South (FTN and FTS, Brown et al. 2013) through the GSP collaboration. The FLOYDS spectra have wavelength coverage of ∼3200–10000 Å with a resolu-tion of ∼18 Å. The spectra were reduced with the `floydsspec`[1] pipeline, using standard reduction techniques. The reduced spectra were scaled to the $BgVRI$ photometry of the corresponding epoch using a linear fit. The log of spectroscopic observations of SN 2019nyk is given in Table 2. Additionally, the classification spectrum obtained by the ePESSTO+ team was downloaded directly from the Transient Name Server[2] (TNS).

**Table 1.** Magnitudes of SN 2019nyk in $UBVgri$ filters. The $UBV$-band photometry are given in Vega magnitudes and $gri$-band data are presented in AB magnitudes.

| Date (yyyy-mm-dd) | JD | Phase[a] (d) | $U$ (mag) | $B$ (mag) | $V$ (mag) | $g$ (mag) | $r$ (mag) | $i$ (mag) |
|---|---|---|---|---|---|---|---|---|
| 2019-08-22 | 2458717.9 | 4.0 | 16.110 ± 0.019 | 16.850 ± 0.016 | 17.006 ± 0.043 | 16.755 ± 0.010 | 17.051 ± 0.026 | 17.333 ± 0.018 |
| 2019-08-23 | 2458718.9 | 5.0 | 16.144 ± 0.021 | 16.865 ± 0.017 | 16.932 ± 0.017 | 16.767 ± 0.013 | 17.084 ± 0.018 | 17.307 ± 0.023 |
| 2019-08-24 | 2458719.9 | 6.0 | 16.365 ± 0.044 | – | 16.957 ± 0.014 | 16.768 ± 0.017 | 17.033 ± 0.050 | 17.219 ± 0.030 |
| 2019-08-26 | 2458721.6 | 7.7 | 16.294 ± 0.019 | 16.916 ± 0.013 | 16.899 ± 0.013 | 16.775 ± 0.006 | 16.977 ± 0.011 | 17.116 ± 0.016 |
| 2019-08-29 | 2458724.7 | 10.8 | 16.543 ± 0.015 | 16.999 ± 0.009 | 16.911 ± 0.010 | 16.850 ± 0.005 | 16.957 ± 0.006 | 17.057 ± 0.008 |
| 2019-08-31 | 2458726.8 | 12.9 | 16.792 ± 0.025 | 17.134 ± 0.012 | 16.981 ± 0.012 | 16.954 ± 0.006 | 16.968 ± 0.008 | 17.077 ± 0.009 |
| 2019-09-02 | 2458728.6 | 14.7 | 16.745 ± 0.020 | 17.121 ± 0.012 | 16.945 ± 0.013 | 16.928 ± 0.005 | 16.950 ± 0.007 | 17.027 ± 0.011 |
| 2019-09-03 | 2458729.8 | 15.9 | 17.047 ± 0.018 | 17.163 ± 0.011 | 17.034 ± 0.012 | 16.984 ± 0.005 | 17.004 ± 0.006 | 17.095 ± 0.010 |
| 2019-09-05 | 2458732.1 | 18.2 | 17.045 ± 0.021 | 17.272 ± 0.012 | 17.069 ± 0.012 | 17.057 ± 0.005 | 17.018 ± 0.007 | 17.094 ± 0.010 |
| 2019-09-09 | 2458736.2 | 22.3 | 17.558 ± 0.028 | 17.479 ± 0.015 | 17.159 ± 0.014 | 17.189 ± 0.007 | 17.074 ± 0.008 | 17.172 ± 0.011 |
| 2019-09-13 | 2458739.7 | 25.8 | 17.561 ± 0.278 | 17.679 ± 0.096 | 17.311 ± 0.068 | – | 17.111 ± 0.030 | 17.258 ± 0.031 |
| 2019-09-17 | 2458743.8 | 29.9 | – | – | – | – | 17.223 ± 0.031 | 17.334 ± 0.045 |
| 2019-09-21 | 2458747.8 | 33.9 | 19.179 ± 0.058 | 18.350 ± 0.019 | 17.577 ± 0.015 | 17.936 ± 0.009 | 17.336 ± 0.009 | 17.400 ± 0.014 |
| 2019-09-25 | 2458752.3 | 38.4 | 19.479 ± 0.088 | 18.631 ± 0.022 | 17.636 ± 0.019 | 18.104 ± 0.015 | 17.374 ± 0.015 | 17.435 ± 0.021 |
| 2019-10-04 | 2458761.3 | 47.4 | 19.901 ± 0.140 | 19.037 ± 0.036 | 17.876 ± 0.018 | 18.442 ± 0.015 | 17.510 ± 0.011 | 17.570 ± 0.015 |
| 2019-10-06 | 2458762.7 | 48.8 | 20.371 ± 0.148 | 19.105 ± 0.033 | 17.975 ± 0.022 | 18.551 ± 0.013 | 17.578 ± 0.011 | 17.650 ± 0.016 |
| 2019-10-10 | 2458766.7 | 52.8 | – | 19.186 ± 0.097 | 18.020 ± 0.043 | 18.622 ± 0.036 | 17.580 ± 0.023 | 17.611 ± 0.031 |
| 2019-10-13 | 2458770.4 | 56.5 | – | 19.333 ± 0.046 | 18.120 ± 0.037 | 18.737 ± 0.036 | 17.562 ± 0.023 | – |
| 2019-10-20 | 2458776.7 | 62.8 | – | 19.573 ± 0.030 | 18.289 ± 0.018 | 18.887 ± 0.019 | 17.754 ± 0.011 | 17.850 ± 0.018 |
| 2019-10-26 | 2458783.1 | 69.2 | – | 20.012 ± 0.060 | 18.593 ± 0.027 | – | – | 18.019 ± 0.031 |
| 2019-11-01 | 2458789.1 | 75.2 | – | 20.175 ± 0.095 | 18.728 ± 0.053 | 19.438 ± 0.033 | 18.003 ± 0.023 | 18.090 ± 0.043 |
| 2019-11-06 | 2458794.3 | 80.4 | – | – | 19.116 ± 0.037 | 20.166 ± 0.043 | 18.528 ± 0.048 | – |
| 2019-11-10 | 2458798.0 | 84.1 | – | 20.875 ± 0.298 | 19.631 ± 0.163 | 20.259 ± 0.114 | 18.718 ± 0.094 | 18.822 ± 0.120 |
| 2019-11-11 | 2458798.6 | 84.7 | – | 21.432 ± 0.484 | 19.587 ± 0.113 | 20.769 ± 0.149 | 18.857 ± 0.060 | 18.965 ± 0.087 |
| 2019-11-13 | 2458800.6 | 86.6 | – | 21.077 ± 0.317 | 20.007 ± 0.117 | 20.912 ± 0.130 | 19.210 ± 0.043 | 19.173 ± 0.050 |
| 2019-11-14 | 2458802.4 | 88.5 | – | 21.690 ± 0.248 | 20.285 ± 0.126 | 21.052 ± 0.143 | 19.265 ± 0.037 | 19.326 ± 0.048 |
| 2019-11-15 | 2458803.4 | 89.5 | – | 21.757 ± 0.428 | 20.072 ± 0.096 | 21.002 ± 0.115 | 19.320 ± 0.054 | 19.385 ± 0.060 |
| 2019-11-18 | 2458805.6 | 91.7 | – | 21.807 ± 0.174 | 20.418 ± 0.083 | 21.329 ± 0.073 | 19.405 ± 0.026 | 19.554 ± 0.057 |
| 2019-11-21 | 2458809.4 | 95.5 | – | – | 20.384 ± 0.068 | 21.421 ± 0.066 | 19.423 ± 0.033 | 19.498 ± 0.067 |
| 2019-11-23 | 2458810.6 | 96.7 | – | – | 20.362 ± 0.080 | 21.435 ± 0.083 | 19.509 ± 0.031 | 19.578 ± 0.063 |
| 2019-11-24 | 2458812.4 | 98.5 | – | – | 20.400 ± 0.074 | 21.347 ± 0.076 | 19.428 ± 0.030 | 19.614 ± 0.064 |
| 2019-11-26 | 2458814.3 | 100.4 | – | – | 20.379 ± 0.067 | 21.369 ± 0.070 | 19.498 ± 0.030 | 19.643 ± 0.063 |
| 2019-11-28 | 2458816.0 | 102.1 | – | – | 20.490 ± 0.117 | 21.346 ± 0.149 | 19.655 ± 0.199 | 19.763 ± 0.226 |
| 2019-12-01 | 2458818.6 | 104.7 | – | – | 20.660 ± 0.196 | 21.374 ± 0.366 | 19.487 ± 0.177 | 19.753 ± 0.298 |
| 2019-12-03 | 2458820.7 | 106.8 | – | – | 20.503 ± 0.063 | 21.371 ± 0.081 | 19.577 ± 0.027 | 19.665 ± 0.042 |
| 2019-12-14 | 2458832.3 | 118.4 | – | – | 20.662 ± 0.108 | – | 19.903 ± 0.093 | – |
| 2019-12-25 | 2458843.3 | 129.4 | – | – | 20.792 ± 0.071 | – | 19.742 ± 0.027 | 19.901 ± 0.061 |
| 2020-01-06 | 2458854.6 | 140.7 | – | – | 20.882 ± 0.143 | – | 19.721 ± 0.033 | 19.951 ± 0.035 |

[a] since the explosion epoch $t_0 =$ JD 2458713.9

## 4. Light curve analysis

The $UBgVri$ LCs of SN 2019nyk along with the ZTF-$g$ and $r$ bands photometric points are shown in Fig. 2, where the ZTF-$g$ and $r$ band magnitudes are shifted by 0.06 and 0.11 mag, respectively. The LCs attain a peak absolute magnitude of −18.25±0.17, −18.09±0.17, −18.02±0.17, and −17.92±0.17 mag in $g$, $V$, $r$ and $i$ bands at 6.1±2.2, 8.6±2.5, 11.3±2.9, and 15.2±2.5 day from explosion, respectively. The decay rates of Type II SNe are often characterised by two slopes: an initial steep slope, $s1$, that immediately follows the peak of the LC and a smaller decay rate, $s2$, that precedes the end of the plateau (Anderson et al. 2014). However, not all H-rich Type II SNe exhibit distinct slopes during the plateau phase (Valenti et al. 2016), as in the case for SN 2019nyk. The decline rate after peak in $V$ band is 2.84±0.03 mag (100 d)$^{-1}$. The transition from the recombination phase to the radioactive tail phase is marked by a drop in magnitude and is referred to as the transition phase. The decline in the radioactive tail phase is denoted as $s3$. In the $g$ and $V$ band LCs, a brief 'plateau tail phase' becomes evident, spanning from 92 to 107 d in the $g$ band and 92 to 103 d in the $V$ band. This plateau is also observed in the $BVRI$ LCs of SN 1999em (Elmhamdi et al. 2003) and is suggested to arise from the radiation emitted by the inner, warmer layers of the ejecta, contributing to an excess in luminosity during the radioactive tail phase (Utrobin 2007). We report the slopes at different phases of the LC in all the observed bands of SN 2019nyk in Table 3.

---

[1] https://www.authorea.com/users/598/articles/6566
[2] https://www.wis-tns.org/object/2019nyk

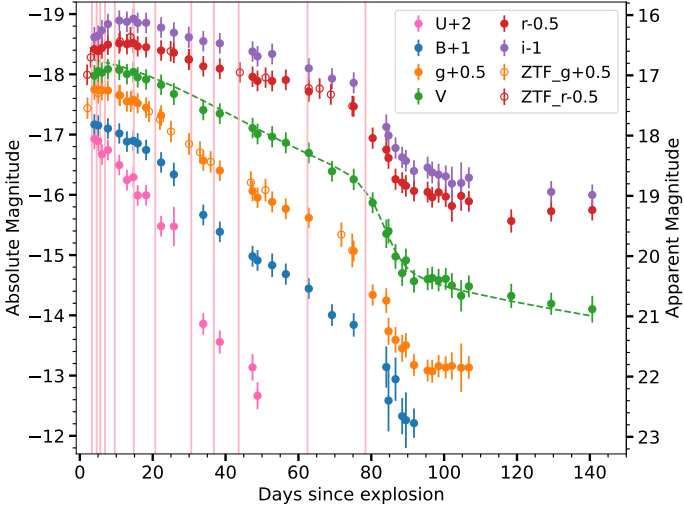

**Fig. 2.** Absolute and apparent $UBgVri$ LCs of SN 2019nyk are shown, offset by values given in the legend. An additional shift of 0.06 and 0.11 mag are applied to the ZTF-$g$ and $r$ band magnitudes, respectively, to match the LCO $g$ and $r$ bands magnitudes. The LCO and shifted ZTF data are shown with filled symbols and open symbols, respectively. The error bars correspond to the error in the absolute magnitudes. The vertical lines denote the epochs at which a spectrum of SN 2019nyk was obtained. The dashed line shows the parametric fit to the $V$ band LC (Olivares et al. 2010).

**Table 2.** Log of spectroscopic observations.

| UT Date (yyyy-mm-dd) | JD | Phase[a] (d) | Telescope | Exposure time (s) |
|---|---|---|---|---|
| 2019-08-22.5 | 2458718.0 | 4.1 | FTN | 2700 |
| 2019-08-23.6 | 2458719.0 | 5.1 | FTN | 2700 |
| 2019-08-24.8 | 2458720.3 | 6.4 | FTS | 2700 |
| 2019-08-27.6 | 2458723.0 | 9.1 | FTS | 2700 |
| 2019-09-01.7 | 2458728.2 | 14.3 | FTS | 2700 |
| 2019-09-07.6 | 2458734.1 | 20.2 | FTS | 2700 |
| 2019-09-17.5 | 2458744.0 | 30.1 | FTN | 2700 |
| 2019-09-23.7 | 2458750.2 | 36.3 | FTS | 2700 |
| 2019-09-30.5 | 2458757.0 | 43.1 | FTN | 2700 |
| 2019-10-19.4 | 2458775.9 | 62.0 | FTN | 3600 |
| 2019-11-04.4 | 2458791.8 | 77.9 | FTN | 3600 |

[a] since the explosion epoch $t_0$ = JD 2458713.9

We use Equation 4 from Olivares et al. (2010) to fit the $V$ band LC in order to estimate the free parameters pertaining to the transition phase of the LC: the drop in magnitude during the transition phase, $a_0$, the duration of the transition phase, $w_0$, and the middle of the transition phase, $t_{PT}$. We estimate a 1.4 ($\pm$ 0.2) mag fall in magnitude ($a_0$) after the recombination phase, in a time span of 17 days ($\pm$ 5), which is roughly 6 $\times$ $w_0$, with a decline rate of 13.7 mag $(100\,d)^{-1}$. The $t_{PT}$ parameter, which denotes the transition point from the recombination to the tail phase, is $83.2 \pm 2.1$ d.

We constructed a sample of SNe II to use as a reference for analysing the LC and spectroscopic characteristics of SN 2019nyk. Specifically, for our comparative dataset, we have assembled a set of SNe with recombination phase decline rates exceeding 1 mag $(100d)^{-1}$. The comparison sample encompasses SN 2013by, which exhibited late-time intermediate interaction signatures, and seven Type II SNe that display observ-

**Table 3.** Light curve slopes in different phases.

| Band | $t_{start}$(d) | $t_{stop}$(d) | slope (mag/100d) | Comment |
|---|---|---|---|---|
| U | 4.5 | 26.3 | 7.2±0.8 | s1 |
| B | 4.5 | 18.7 | 2.5±0.1 | s1 |
|   | 22.8 | 75.7 | 4.82±0.09 | s2 |
| g | 2.6 | 8.2 | −4.3±0.7 | rise |
|   | 11.3 | 80.9 | 4.42±0.03 | s1=s2 |
|   | 84.6 | 89.0 | 15±3 | fall |
|   | 92.2 | 107.2 | −0.06±1.25 | s3-plateau |
| V | 4.5 | 11.3 | −1.2±0.5 | rise |
|   | 13.4 | 75.7 | 2.84±0.03 | s1=s2 |
|   | 80.9 | 87.2 | 14±2 | fall |
|   | 92.0 | 102.6 | 0.47±1.09 | s3-plateau |
|   | 105.2 | 141.2 | 0.86±0.48 | s3 |
| r | 2.5 | 8.2 | −8.1±0.6 | rise |
|   | 11.3 | 75.7 | 1.7±0.01 | s1=s2 |
|   | 80.9 | 87.2 | 10±1 | fall |
|   | 89.0 | 141.2 | 0.95±0.07 | s3 |
| i | 4.5 | 8.2 | −6.2±0.4 | rise |
|   | 11.3 | 69.7 | 1.55±0.02 | s1=s2 |
|   | 75.7 | 87.2 | 9.2±0.4 | fall |
|   | 89.0 | 141.2 | 1.13±0.09 | s3 |

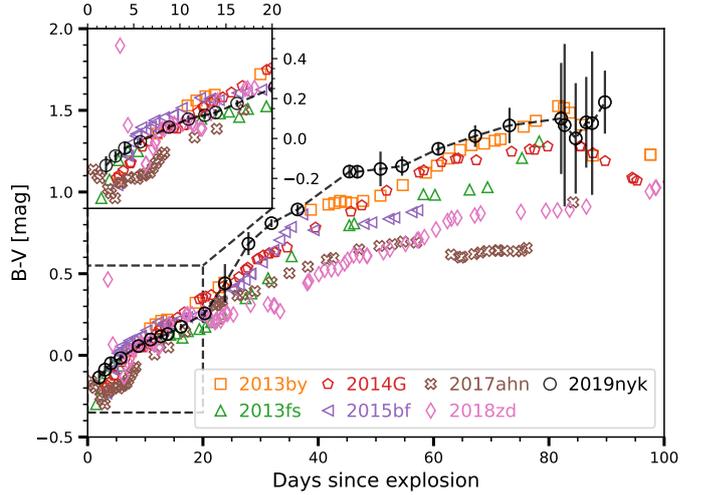

**Fig. 3.** (B − V) colour curve of SN 2019nyk compared to those of the comparison sample SNe. Prior to estimating the colour, LOESS regression was applied to smooth out the B and V band LCs of all SNe. The inset plot shows a zoom-in of the first 20 days of the evolution, which shows the colour evolution of SNe 2017ahn and 2018zd to bluer colours before transitioning to redder colours.

able 'IIn-like' features in their early spectra: SN 2013fs, 2014G, 2015bf, 2017ahn, 2018zd, 2020pni and 2023ixf. In Table 4, we provide the parameters of the comparison sample utilised in this study, along with the corresponding references.

### 4.1. Colour and temperature evolution

The (B − V) colour evolution of SN 2019nyk is shown along with that of the comparison sample in Fig. 3. We applied the non-parameteric smoother, LOESS-locally weighted running line smoother, to fit the B and V band LCs before estimating the colours for all the SNe. As is apparent from the figure, the (B − V) colour evolution of SN 2019nyk is typical of SNe II. The (B − V) colour of SN 2019nyk becomes redder rapidly in the first 10 days when the temperature drops quickly following

Table 4. Type II SNe comparison sample.

| SN | Host galaxy | Explosion Epoch (MJD) | Redshift | Distance modulus (mag) | E(B−V)$_{host}$ (mag) | References |
|---|---|---|---|---|---|---|
| 2013by | ESO138-G010 | 56401.6 ± 3.6 | 0.00382 | 30.46 ± 0.29 | – | 1 |
| 2013fs | NGC 7610 | 56570.8 ± 0.5 | 0.01185 | 33.32 ± 0.18 | – | 4, 5 |
| 2014G | NGC 3448 | 56669.3 ± 0.8 | 0.00450 | 31.96 ± 0.14 | 0.268 ± 0.046 | 2, 3 |
| 2015bf | NGC 7653 | 57367.3$^{+1.2}_{-2.4}$ | 0.013654 | 33.89 ± 0.05 | 0.148 | 6 |
| 2017ahn | NGC 3318 | 57791.8 ± 0.5 | 0.00925 | 32.84 ± 0.10 | 0.233 ± 0.148 | 7 |
| 2018zd | NGC 2146 | 58178.4$^{+0.2}_{-0.5}$ | 0.002979 | 31.32 ± 0.53 | 0.17 ± 0.05 | 8, 9, 10 |
| 2020pni | UGC 09684 | 59045.8 ± 0.1 | 0.01665 | 34.34 ± 0.15 | 0.080 ± 0.010 | 11 |
| 2023ixf | M101 | 60082.7 ± 0.01 | 0.000804 | 29.19 ± 0.04 | 0.038 ± 0.009 | 12, 13, 14, 15, 16, 17 |
| 2019nyk | HIPASS J0015-08 | 58713.9 ± 2.0 | 0.02100 | 34.88 ± 0.17 | – | This work |

References: [1] Valenti et al. (2015), [2] Bose et al. (2016), [3] Terreran et al. (2016), [4] Yaron et al. (2017), [5] Bullivant et al. (2018), [6] Lin et al. (2021), [7] Tartaglia et al. (2021), [8] Zhang et al. (2020), [9] Hiramatsu et al. (2021), [10] Callis et al. (2021), [11] Terreran et al. (2022), [12] Jencson et al. (2023), [13] Jacobson-Galán et al. (2023), [14] Hosseinzadeh et al. (2023), [15] Bostroem et al. (2023), [16] Teja et al. (2023), [17] Hiramatsu et al. (2023).

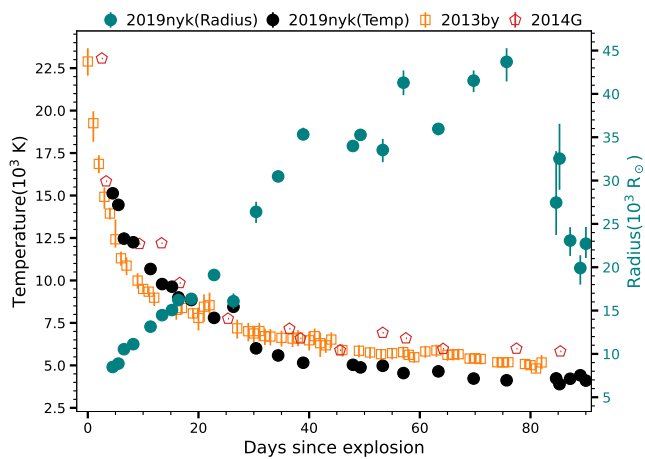

Fig. 4. Temperature and radius evolution of SN 2019nyk. The temperature evolution of SNe 2013by and 2014G are also shown.

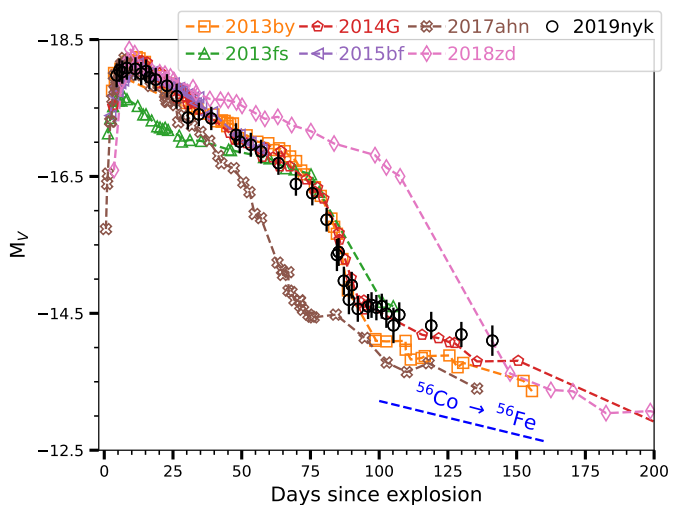

Fig. 5. Comparison of absolute V band LCs of SN 2019nyk with those of the comparison sample. The magnitudes are corrected for distance and reddening as listed in Table 5. The radioactive decay line assuming full trapping of photons is shown with a dashed line.

a power law in time. Thereafter, we see a slower evolution until 20 days, followed by a rapid change in colour, increasing from 0.3 to 0.8 mag between 20 and 50 days. The (B − V) colour of SN 2019nyk, in contrast to the SNe from the comparison sample, exhibits similar colour during the early phases (<20 d), except for SNe 2017ahn and 2018zd, which show an evolution towards bluer colours before transitioning to redder colours. Beyond 30 days, the colour of SN 2019nyk closely matches that of SNe 2013by and 2014G, but it is redder than the colours of other SNe in the comparison sample.

Fig. 4 shows the photospheric temperature and radius evolution of SN 2019nyk, estimated by fitting blackbody function to the SEDs generated from photometry using the LC fitting code of Hosseinzadeh & Gomez (2020). For comparison, we have also plotted the temperature evolution of SNe 2013by and 2014G. In the case of SN 2013by, early interaction signatures were not detected, while for SN 2014G, there were detections. The temperature of all the SNe decreases with time as the ejecta expands and cools. At the early phases (<10 d), the temperature of SN 2019nyk is similar to SN 2014G. From the beginning of the recombination phase in SN 2019nyk, the temperature decreases faster than SNe 2013by and 2014G. The radius increases initially up to 75 d, which corresponds to the end of the recombination phase, then falls off rapidly as the ejecta becomes optically thin.

### 4.2. Absolute magnitude and $^{56}$Ni mass

The V band absolute magnitude evolution of SN 2019nyk is shown along with the comparison sample in Fig 5. The evolution of SN 2019nyk closely resembles that of SNe 2013by and 2014G, exhibiting similar peak magnitudes (refer to Table 5). However, the drop in luminosity after the recombination phase is more significant in SN 2013by compared to SN 2019nyk. Moreover, SN 2013by does not exhibit the 'plateau tail phase' feature observed in the case of SN 2019nyk. Although the tail luminosity of SN 2019nyk is quite similar to that of SN 2014G, the decline rate during the tail phase for SN 2019nyk (1.20 ± 0.20 mag 100d$^{-1}$) is shallower than that of SN 2014G (1.68 ± 0.18 mag 100d$^{-1}$) and other fast decliners in the comparison sample (see Table 5).

The amount of $^{56}$Ni synthesised during the explosion can be inferred using the radioactive tail photometry. During this phase, the LC is powered by the energy from the radioactive decay of $^{56}$Co to $^{56}$Fe, generating gamma rays which subsequently interact with the ejecta, leading to the production of optical light. To estimate the mass of $^{56}$Ni, we employ three distinct methods, each elaborated in detail as follows.

**Table 5.** Parameters of comparison sample in the *V* band.

| SN | s1 (mag/100d) | s2 (mag/100d) | s3 (mag/100d) | $t_V^{peak}$ (d) | $M_V^{peak}$ (mag) | $M_V^{50d}$ (mag) | $^{56}$Ni ($M_\odot$) | $t_{PT}$ (d) |
|---|---|---|---|---|---|---|---|---|
| 2013by | 3.43 ± 0.02 | 1.98 ± 0.04 | 1.20 ± 0.06 | 12.0 ± 2.0 | −18.05 ± 0.15 | −17.40 ± 0.44 | 0.0327 ± 0.0117 | 84.0 ± 3.6 |
| 2013fs | 4.47 ± 0.26 | 1.25 ± 0.03 | – | 7.6 ± 1.0 | −17.66 ± 0.05 | −16.74 ± 0.21 | 0.0490 ± 0.0121 | 84.6 ± 3.0 |
| 2014G | – | 2.87 ± 0.05 | 1.68 ± 0.18 | 10.8 ± 2.6 | −18.05 ± 0.84 | −17.18 ± 0.47 | 0.0726 ± 0.0130 | 85.9 ± 1.0 |
| 2015bf | – | 2.71 ± 0.08 | – | 11.3 ± 3.3 | −17.95 ± 0.06 | −17.25 ± 0.08 | – | – |
| 2017ahn | – | 4.02 ± 0.09 | 1.90 ± 0.07 | 7.5 ± 0.6 | −18.40 ± 0.50 | −16.75 ± 0.46 | 0.0506 ± 0.0157 | $59.3^{+1.5}_{-1.9}$ |
| 2018zd | – | 1.60 ± 0.01 | 0.88 ± 0.02 | 9.0 ± 0.9 | −18.40 ± 0.60 | −17.79 ± 0.55 | 0.0300 ± 0.0161 | $117.9^{+2.0}_{-2.2}$ |
| 2020pni | – | 3.90 ± 0.60 | – | 10.2 ± 2.9 | −18.21 ± 0.08 | −16.89 ± 0.25 | – | – |
| 2019nyk | – | 2.84 ± 0.03 | 1.20 ± 0.20 | 8.6 ± 2.5 | −18.09 ± 0.17 | −16.95 ± 0.17 | 0.050 ± 0.010 | 83.2 ± 2.1 |

First, we employ the bolometric correction (BC) method to estimate the $^{56}$Ni mass from the tail phase photometry. We use the magnitudes in the *V* band radioactive tail at three epochs, that is, 118.9, 129.9 and 141.2 d from explosion and the BC from Hamuy (2003), to obtain an average bolometric luminosity $L_{bol,tail}$ = 1.98 ± 0.19 × 10$^{41}$ erg s$^{-1}$. The resulting $^{56}$Ni mass is 0.047±0.004 $M_\odot$. We also estimate the $^{56}$Ni mass, using the sloan *i* band magnitudes at 129.9 and 141.2 d and the BC given in Table 9 of Rodríguez et al. (2021), to be 0.050±0.010 $M_\odot$. Both estimates are consistent within the errors. The $^{56}$Ni mass ejected in SN 2019nyk and that of the comparison sample given in Table 5 are estimated using the method of Rodríguez et al. (2021), to maintain consistency.

Another method to estimate the $^{56}$Ni mass is by assuming that the SN has the same SED as the well-characterised SN 1987A. The $^{56}$Ni mass synthesised in the SN explosion can then be found by scaling the pseudo-bolometric luminosity of SN when it is powered by $^{56}$Co to that of SN 1987A with the following equation (Spiro et al. 2014):

$$M(^{56}\text{Ni}) = 0.075\,M_\odot \times \frac{L_{SN}(t)}{L_{87A}(t)},$$

where M($^{56}$Ni) is the synthesised $^{56}$Ni mass, $L_{SN}(t)$ and $L_{87A}(t)$ are the pseudo-bolometric luminosities of the SNe 2019nyk and 1987A at time t, respectively. Here 't' is the time since explosion. The same filters were used to compute the pseudo-bolometric luminosity of the SNe 2019nyk and 1987A. This scaling can be performed throughout the radioactive tail phase, provided that the slope matches that of SN 1987A, indicating complete trapping of gamma rays. Although for SN 2019nyk, the radioactive tail starts at 92 d, the radioactive tail of SN 1987A starts at 120 d. So, we use the 130 d *Vri* pseudo-bolometric luminosity of SN 2019nyk and obtain pseudo-bolometric luminosity ratio $L_{SN}(130d)/L_{87A}(130d)$ = 0.92 ± 0.02. Using this ratio and the $^{56}$Ni mass of 0.075 ± 0.005 $M_\odot$ estimated for SN 1987A, we estimate M($^{56}$Ni, 2019nyk) = 0.069 ± 0.005 $M_\odot$, which is ∼38 per cent higher than the former estimate.

The third method adopted to estimate the $^{56}$Ni mass is the empirical relation between the *V* band decline rate during the transitional phase (steepness parameter, S = −dM$_V$/dt) and the ejected $^{56}$Ni mass as reported in Elmhamdi et al. (2003) based on 10 SNe II. This work was later improved upon in Singh et al. (2018) using a sample of 39 SNe II. A steepness parameter of 0.31±0.01 is determined for SN 2019nyk using its *V* band LC, which yields a $^{56}$Ni mass of 0.066±0.044 $M_\odot$ using the relation from Singh et al. (2018). This is consistent with the estimates derived from the above two methods, owing to the large associated error. However, Rodríguez et al. (2021) found that this method is not well-suited for measurements of $^{56}$Ni mass of Type II SNe.

The absolute magnitude at 50 d in *V* band is known to be strongly correlated with the synthesised $^{56}$Ni mass for hydrogen-rich Type II SNe, as both these parameters are dependent on the explosion energy. SN 2019nyk conforms to this correlation as shown in Fig. 6. The SNe II sample is obtained from literature sample work (Anderson et al. 2014; Valenti et al. 2016; Rodríguez et al. 2021). Certain SNe from our comparison sample have been highlighted for emphasis. In this representation, the size of the markers corresponds to the steepness of the LC slope (*s2*). Additionally, the colouration of the markers are based on the LC $t_{PT}$ values (except for the comparison sample SNe), with the darkest and lightest copper coloured points representing SNe LC with $t_{PT}$ <80 days and $t_{PT}$ >100, respectively. From this figure, it is apparent that the LCs with steeper slopes have shorter plateau lengths, consistent with findings from prior studies. The steeper and shorter plateau length LCs (fast-declining), however, do not separate from the shallower and longer plateau length LCs (slow-declining) in this parameter space. Consequently, it becomes evident that variations in the mass and density of the hydrogen envelope hold negligible influence over both the synthesised $^{56}$Ni mass and the absolute magnitude at 50 d, and hence, on the explosion energy.

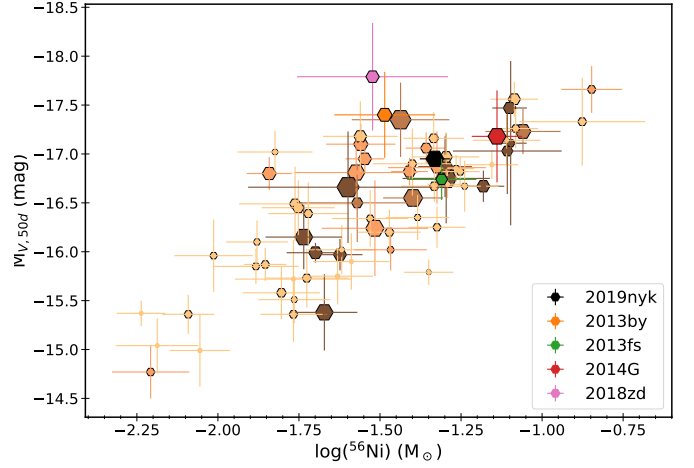

**Fig. 6.** Position of SN 2019nyk on the M$_{V,50d}$ vs $^{56}$Ni mass plot shown alongside other SNe II. The larger marker sizes correspond to the higher values of the LC slopes s2, and the increasing darkness of colour represents SNe with decreasing $t_{PT}$.

## 5. Spectral analysis

### 5.1. Spectral evolution

The earliest spectrum of SN 2019nyk, obtained by ePESSTO+ team at 2.8 d post-explosion, and the first LCO spectrum acquired at 4.1 d post-explosion, are displayed in Fig. 7. The spectra have been continuum-normalised, and key emission lines, in-cluding H α, H β, He II λ4686 and C IV λλ5801, 5811, have been marked for reference.

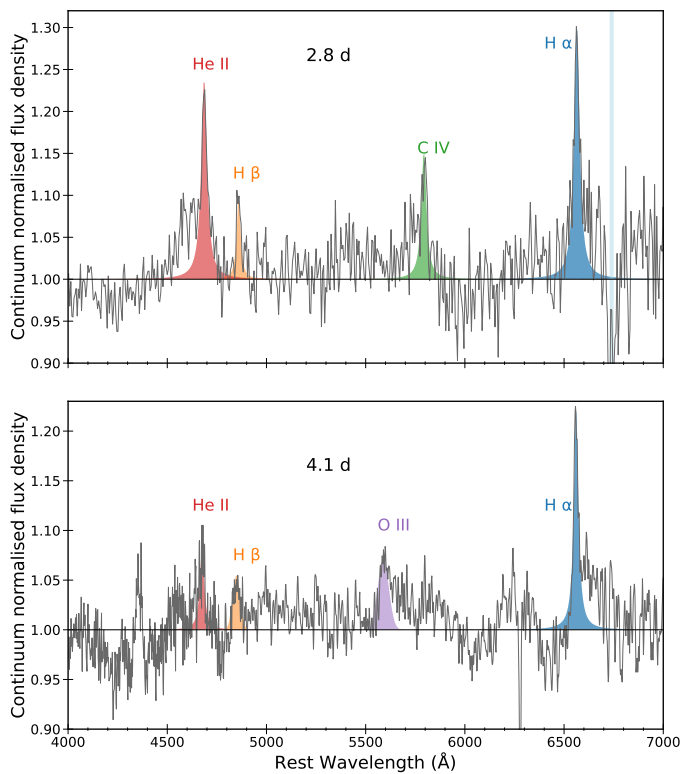

**Fig. 7.** Continuum normalised 2.8 d ePESSTO+ spectrum and 4.1 d LCO spectrum of SN 2019nyk are shown in the upper and lower panels, respectively. The intermediate width emission lines of H α, H β, C IV and He II are marked, which originate from RSG wind. The features are fitted with a Lorentzian profile in order to fit the broad wings.

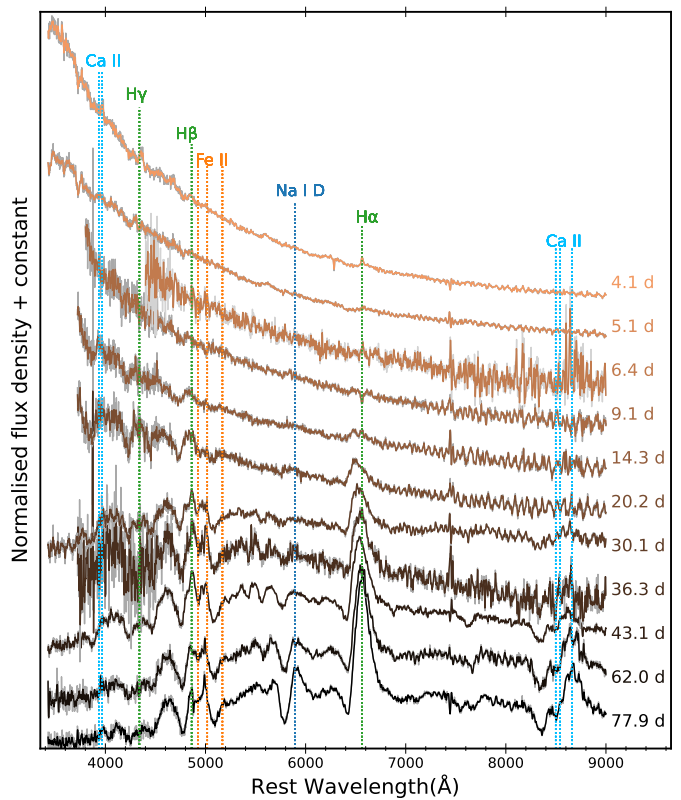

**Fig. 8.** Spectral evolution of SN 2019nyk from 4.1 to 77.9 d is shown. The spectra are corrected for redshift and Galactic extinction.

Numerous studies (e.g. Gal-Yam et al. 2014; Groh 2014; Khazov et al. 2016; Yaron et al. 2017; Bruch et al. 2021, 2023 and references therein) have shown that early spectra (until ∼5 d after the explosion) can be dominated by strong high-ionisation emission lines produced either by the recombination of CSM excited and ionised by the shock-breakout flash or by the recombination of unshocked CSM stimulated by radiation stemming from shocks due to underlying ejecta-CSM interaction. The transient nature of the emission lines in the early spectra of SN 2019nyk indicates that these features originate from a nearby confined CSM, which resulted from increased mass loss occurring months to a few years prior to the explosion.

Given the low resolution of the early spectra of SN 2019nyk, any narrow emission component arising from the CSM (with velocities of the order of 100 km s$^{-1}$) is expected to be entirely washed out. In fact, the emission lines indicated in Fig. 7 likely correspond to intermediate-width wings (velocities close to 1000 km s$^{-1}$), believed to be the result of thermal electron scattering of the narrow-line emissions. This process broadens the narrow-line photons into wings, which can be approximated by a Lorentzian shape (Chugai 2001; Dessart et al. 2009; Huang & Chevalier 2018).

We fit the intermediate width emission features in the 2.8 d spectrum with Lorentzian profiles. The velocity estimated from the FWHM of emission component of He II, H β, C IV and H α are 1110±170, 740±320, 790±210 and 760±120 km s$^{-1}$, respectively. The broad emission feature bluewards of He II most likely arises from C III λλ4647, 4650 and/or N III λλ4634, 4640. This feature has also been reported in the early spectra of SNe 2013fs and 2023ixf. A broad and weak emission feature bluewards of C IV can also be seen in the 2.8 d spectrum.

While the 4.1 d spectrum retains the intermediate-width H α emission (FWHM velocity: 640 km s$^{-1}$) and possibly He II (FWHM velocity: 840 km s$^{-1}$), the H β line broadens significantly (FWHM velocity: 2800 km s$^{-1}$). Additionally, an emission component on the red wing of H α exhibits a FWHM velocity of 2100 km s$^{-1}$, comparable to the FWHM of the broad H β feature. Notably, a new emission feature appears at 5590 Å. If this corresponds to O III λ5592.37, then its FWHM velocity is 3510 km s$^{-1}$. Such lines with widths of a few 1000 km s$^{-1}$ are believed to arise from the 'cold dense shell' formed from the piling up of the cool material between the shocked ejecta and the shocked CSM (Chugai et al. 2004).

The spectral evolution of SN 2019nyk from 4.1 to 77.9 d is shown in Fig. 8. Until 14.3 d, the spectra shows a blue continuum, with emerging H Balmer P Cygni profiles. A boxy emission component of the H α P Cygni profile along with a weak absorption component is seen in the 20.2 d spectrum, and was also observed in the spectra of SNe 2007od, 2016esw and 2016gfy (Inserra et al. 2011; de Jaeger et al. 2018; Singh et al. 2019). The existence of this feature serves as an indicator of ongoing interaction, with its origin attributed to the absorption of shock energy by the dense shell formed through the accumulation of swept-up CSM and decelerated ejecta material (Dessart et al. 2016). Metal lines, such as, Fe II λλ5018,5169, Na ID, Ca II NIR becomes conspicuous from the 30.1 d spectrum.

In order to identify the elements contributing to the photospheric spectra of SN 2019nyk, we used SYNAPPS (Thomas et al. 2011; Thomas 2013) to produce a synthetic spectrum that resembles the observed spectrum at 43.1 d. SYNAPPS is an open-source spectrum synthesis code, an automated version of SYN++, wherein an optimisation method is employed to output the best-fitted parameters. The 43.1 d spectrum is chosen for modelling as it has a relatively high signal-to-noise ratio, and there are visibly many strong features. The top panel of Fig. 9 compares the synthetic and observed spectra produced with SYNAPPS, and the seven lower panels show the SYN++ models by activating one species at a time in order to show the contribution of each of the seven ions used for the modelling. The modelling shows that the blue part of the spectrum (< 5200 Å) is dominated by Fe II, Sc II and Ti II lines. The dip at 7600 Å is a blend of telluric features, and hence not reproduced by the model.

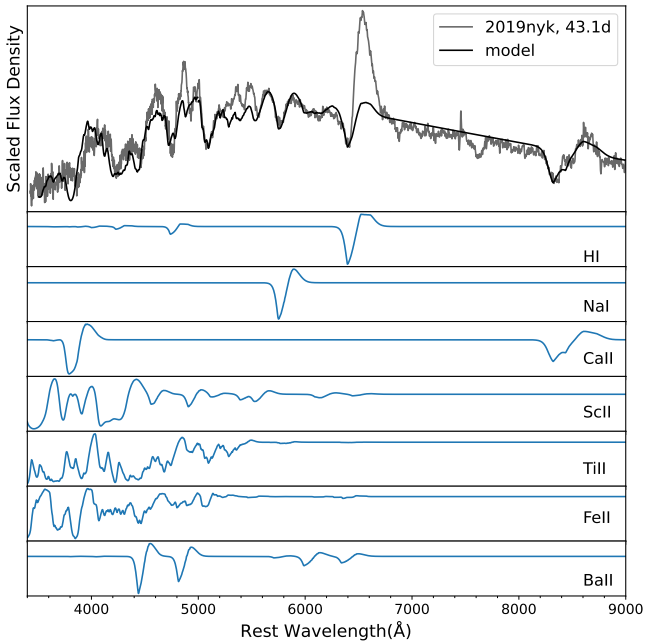

**Fig. 9.** SYNAPPS model for the 43.1 d spectrum of SN 2019nyk is shown in the top panel along with the observed spectrum. The seven bottom panels depict the specific contributions of each of the seven ions in the top panel's synthetic spectrum. The models in the bottom panels are generated using SYN++ by independently activating each of the seven ions.

The expansion velocities and the associated uncertainties of the prominent features in the spectra, such as H $\alpha$, H $\beta$, Fe II, Na ID are estimated from the minimum of their absorption component using the method outlined in Faran et al. (2014). The method basically involves fitting a third order polynomial to the absorption component of the P Cygni profile to estimate the minima and applying relativistic Doppler formula to measure the expansion velocity from the blue-shifted minima. In order to estimate the uncertainty, a noise vector is created by dividing the original spectrum by their smoothed versions. This noise vector is then randomly sampled to draw 1000 new noise vectors, which are multiplied by the smoothed spectrum to generate 1000 new spectra of signal plus noise. Finally, the uncertainty is estimated from the standard deviation of the measurements of minima of 1000 new spectra. The velocity evolution is shown in Fig. 10.

The Fe II $\lambda$5169 velocity does not change much and remains in the range of 4000 - 5000 km s$^{-1}$ within 30 - 80 d. The 'Cachito' feature (Gutiérrez et al. 2017), bluewards of the P Cygni absorption component of H $\alpha$, becomes visible from the 20.2 d spectrum until 36.3 d spectrum. This feature has been identified in 70 SNe in the sample of Gutiérrez et al. (2017) and they concluded that at early times this feature corresponds to Si II, which forms at high temperatures shortly after shock break-out. The 'Cachito' velocity, when interpreted as Si II $\lambda$6355 matches well with the Fe II $\lambda$5169 velocity as shown in Fig. 10. Furthermore, we do not detect any high velocity H $\beta$ feature, hence we conclude that the 'Cachito' is arising from Si II.

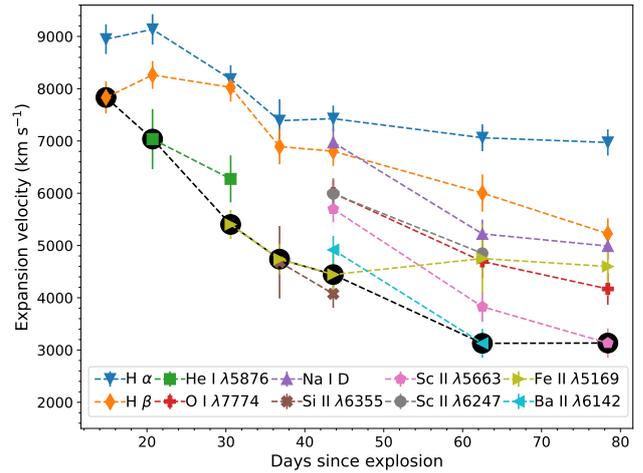

**Fig. 10.** Velocity evolution of the prominent lines in the spectra of SN 2019nyk, where the velocities were estimated from the minima of the absorption profiles. The black dashed line depicts the adopted photospheric velocity evolution for SN 2019nyk and used in Sect. 6, which corresponds to the velocity of the element with the lowest velocity at each epoch.

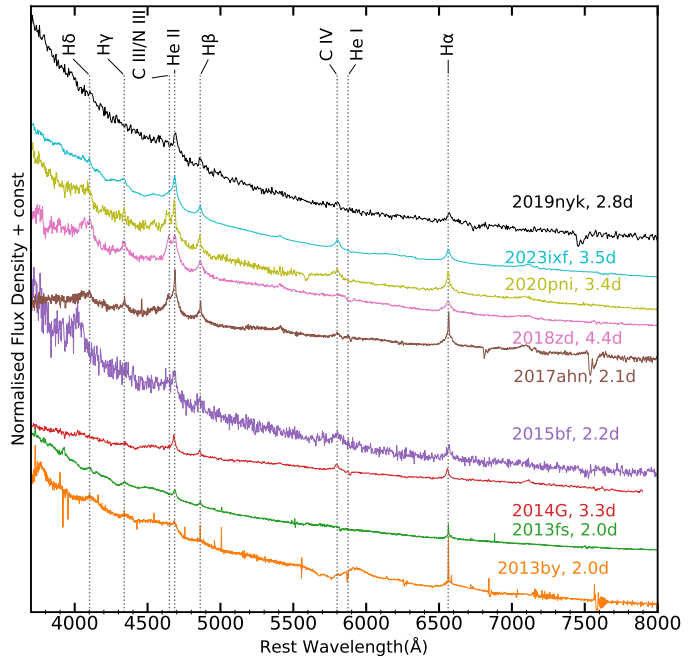

**Fig. 11.** Comparison of early spectrum (2.8 d) of SN 2019nyk with other SNe from the comparison sample at similar epochs.

### 5.2. Comparison with other SNe

In Fig. 11, we present the 2.8 d spectrum of SN 2019nyk together with the spectra of other SNe from the comparison sample. The narrow/intermediate width features of H $\alpha$, H $\beta$, and He II are clearly observed in all SNe. The C IV feature is also present in SNe 2014G, 2017ahn, 2020pni, and 2023ixf, similar to SN 2019nyk, with particular broadening observed in the case of SN 2015bf. Furthermore, the C III/ N III feature appears prominently as narrow emission in SNe 2017ahn, 2018zd, and 2020pni. However, for the remaining SNe, including SN 2019nyk, this feature either appears as a blue-shifted broad emission wing of He II or is entirely absent. In cases with CNO-processed surface abundances, this feature is expected to be dominated by N III, whereas for solar-like surface abundances, C III emission dominates this feature (Boian & Groh 2020).

**Table 6.** Velocity of the intermediate width component in the early time spectra.

| SN | Epoch (d) | H$\alpha$ (km s$^{-1}$) | H$\beta$ (km s$^{-1}$) | He II (km s$^{-1}$) | C IV (km s$^{-1}$) |
|---|---|---|---|---|---|
| 2019nyk | 2.8 | 760 ± 120 | 740 ± 320 | 1110 ± 170 | 790 ± 210 |
| 2023ixf | 3.2 | 780 ± 20 | 840 ± 70 | 760 ± 50 | 1090 ± 50 |
| 2020pni | 3.4 | 540 ± 20 | 740 ± 80 | 710 ± 60 | 500 ± 20 |
| 2014G | 3.3 | 480 ± 20 | 630 ± 90 | 750 ± 60 | 890 ± 70 |
| 2013fs | 2.0 | 730 ± 50 | 630 ± 90 | 600 ± 40 | – |

The velocities estimated for the intermediate-width component of the emission features in the early-time spectra of some comparison sample SNe are estimated by fitting a Lorentzian profile and provided in Table 6. The variations in the FWHM velocities of these features stem from the distinct emitting regions of these species. Highly ionised species tend to form at lower radii, where temperature and opacity are comparatively higher. Consequently, the higher opacity leads to a greater number of electron scattering events, resulting in broader wings for the highly ionised species.

In Fig. 12, we compare the spectra of SNe 2019nyk and 2023ixf observed within the first week after the explosion. The emission line intensity, as well as the FWHM of H$\alpha$ (until 4.1 d) and H$\beta$ (only in the 2.8 d spectrum), are similar in the spectra of SNe 2019nyk and 2023ixf, while the intensity of the He II and C IV features are lower in SN 2019nyk compared to SN 2023ixf. This difference may be attributed to variations in the abundances of these elements on the progenitor surfaces of the respective SNe and/or the temperature of the CSM.

In Fig. 13, the 43.1 d spectrum of SN 2019nyk is compared with similar epoch spectra of the comparison sample. Most of the features in SN 2019nyk spectrum are analogous to those of the comparison sample. The only exception is SN 2017ahn, whose spectrum shows weak absorption features.

In Fig. 14, we present a comparison of the velocity evolution of H$\alpha$, H$\beta$, and Fe II $\lambda$5169 for SN 2019nyk. We juxtapose this with the mean and standard deviation of the velocity evolution calculated for a sample of SN II from Gutiérrez et al. (2017), depicted as a solid grey line and a shaded region, respectively. We also include the velocity profiles of SNe 2013fs, 2014G, 2015bf, 2018zd and 2020pni from our comparison dataset. An increasing velocity trend at early times in H$\alpha$ and H$\beta$ is particularly evident in SNe 2018zd and 2020pni, while to a lesser extent in SN 2013fs. The low velocity at early times is another manifestation of CSM interaction, as the kinetic energy of the ejecta is converted into radiation. After 20 days, SN 2019nyk exhibits H$\alpha$ velocity evolution similar to that of SNe 2013fs, 2014G, 2015bf, and 2020pni. However, before 20 days, SN 2020pni has the lowest H$\alpha$ velocity, and SN 2013fs displays the highest. The H$\alpha$ velocities of SNe 2014G and 2015bf are roughly 1000 km s$^{-1}$ lower than those of SN 2019nyk in the early phases.

Moreover, the H$\beta$ velocity of SN 2014G consistently remains higher than that of SN 2019nyk by at least 1000 km s$^{-1}$.

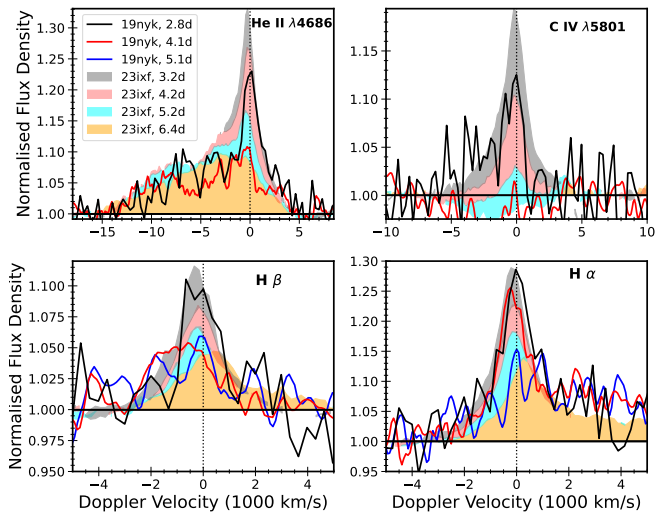

**Fig. 12.** Comparison of the high ionisation emission lines in the early spectra of SNe 2019nyk and 2023ixf. All the spectra of SN 2023ixf used for comparison in this plot were acquired using ALFOSC and Grism 4 (~ 13 Å) mounted on the Nordic Optical Telescope, which is comparable in resolution to the spectra of SN 2019nyk used in this plot (~ 18 Å).

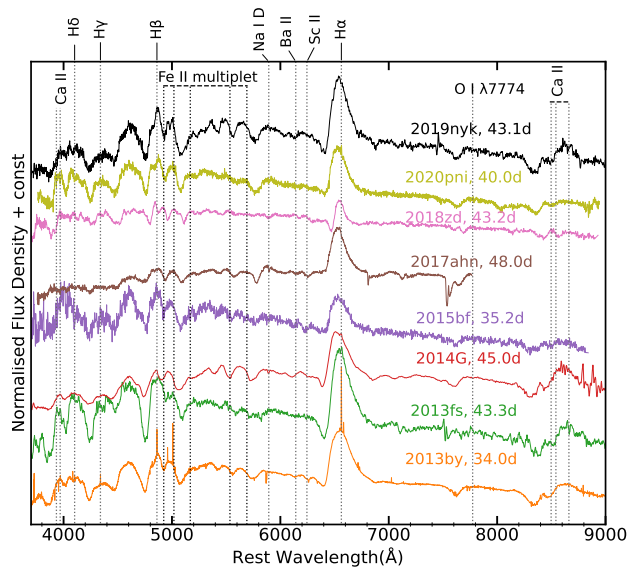

**Fig. 13.** Comparison of 43.1 d spectrum of SN 2019nyk with the nearby epoch spectra of SNe from the comparison sample.

Initially, SN 2013fs has a higher velocity than SN 2019nyk, but after 25 days, they exhibit similar velocities. Concerning Fe II velocity, SN 2019nyk aligns closely with the profiles of SNe 2013fs and 2015bf, albeit remaining 2000 km s$^{-1}$ lower than SNe 2014G and 2020pni until 50 days. However, later in the evolution, they converge due to the flat velocity trend of SN 2019nyk.

### 5.3. Mass-loss rate estimation from the early spectra

We can obtain an order-of-magnitude estimate for the mass-loss rate based on the measured Balmer H$\alpha$ luminosity, following the method outlined in Ofek et al. (2013). Here, we assume that the CSM around the progenitor has a spherical wind density profile of the form $\rho = Kr^{-2}$, where r is the distance from the progenitor and K $\equiv$ $\dot{\mathrm{M}}/(4\pi v_{\mathrm{wind}})$ is the mass-loading parameter (with $\dot{\mathrm{M}}$ being the mass-loss rate and $v_{\mathrm{wind}}$ the wind expansion velocity). Based on the H$\alpha$ line flux on 2.8 and 4.1 d,

$$f_{\mathrm{H}\alpha,2.8d} = (3.7 \pm 0.7) \times 10^{-14}\,\mathrm{erg\,s^{-1}\,cm^{-2}}$$

and

$$f_{\mathrm{H}\alpha,4.1d} = (2.1 \pm 0.3) \times 10^{-14}\,\mathrm{erg\,s^{-1}\,cm^{-2}}$$

as measured from the reddening-corrected spectra, the H$\alpha$ luminosity are $L_{\mathrm{H}\alpha,2.8d} = f_{\mathrm{H}\alpha,2.8d} 4\pi d^2 = (4.0 \pm 1.0) \times 10^{40}\,\mathrm{erg\,s^{-1}}$ and $L_{\mathrm{H}\alpha,4.1d} = f_{\mathrm{H}\alpha,4.1d} 4\pi d^2 = (2.3 \pm 0.4) \times 10^{40}\,\mathrm{erg\,s^{-1}}$, where d = 94.7 ± 7.4 Mpc is the luminosity distance to the SN.

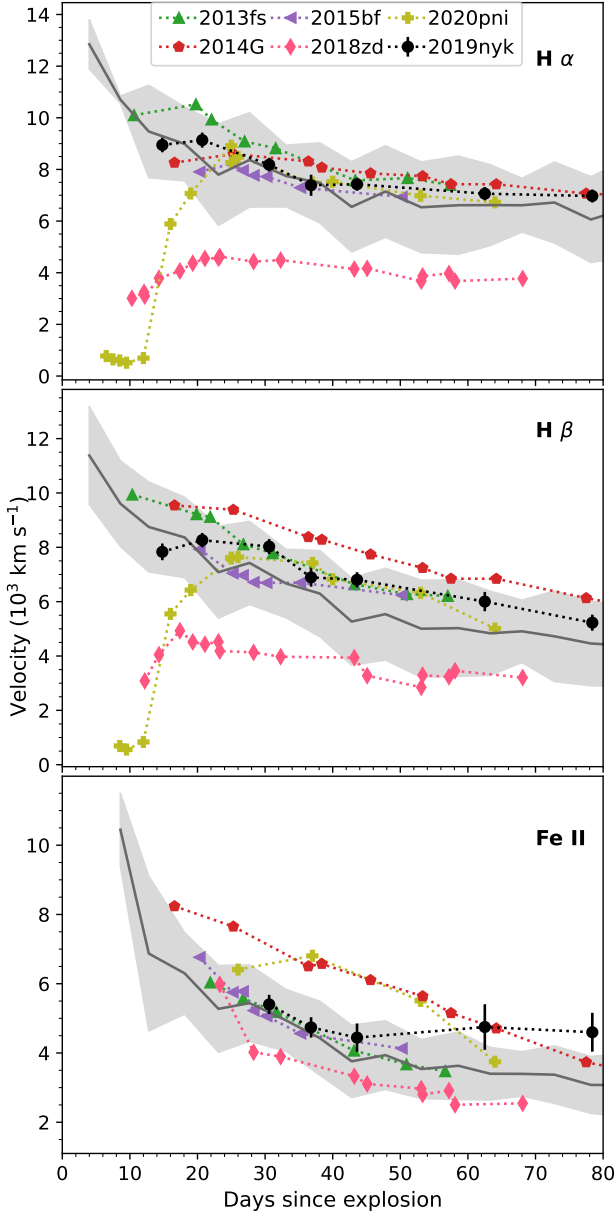

**Fig. 14.** Expansion velocity of SN 2019nyk is shown along with that of the comparison sample SNe and the mean velocity (grey) of 122 IIP/IIL SNe (Gutiérrez et al. 2017) for H$\alpha$ (top panel), H$\beta$ (middle panel), and Fe II ($\lambda$5169; bottom panel). The light grey regions represent the standard deviations of the mean velocities of the sample.

According to Eq. 4 in Ofek et al. (2013), a relation can be obtained between the mass-loading parameter K, the H$\alpha$ luminosity ($L_{\mathrm{H}\alpha}$), the radius (r) and the temperature ($T_{\mathrm{eff}}$):

$$\begin{aligned} \mathrm{K} &\gtrsim \left(\frac{\langle\mu_p\rangle m_p^2 L_{\mathrm{H}\alpha} r}{4\pi h \nu_H \alpha_H^{\mathrm{eff}}}\right)^{1/2} \\ &\approx 7.1 \times 10^{15} \times \left(\frac{L_{\mathrm{H}\alpha}}{10^{41}\,\mathrm{erg\,s^{-1}}}\right)^{1/2} \times \left(\frac{r^2}{(r-r_0)10^{15}\,\mathrm{cm}}\right)^{1/2} \\ &\quad \times \left(\frac{T_{\mathrm{eff}}}{10^4\,\mathrm{K}}\right)^{0.89/2}\,\mathrm{g\,cm^{-1}} \end{aligned} \quad (1)$$

$\langle\mu_p\rangle$ = 0.6 is the adopted mean molecular weight, $\nu_H$ is the H$\alpha$ line frequency (4.57 × 10$^{14}$ Hz for H$\alpha$), $r_0$ is the inner radius of the wind, and the H$\alpha$ effective recombination coefficient,

$$\alpha_H^{\mathrm{eff}} \approx 8.7 \times 10^{-14} (T_{\mathrm{eff}}/10^4\,\mathrm{K})^{-0.89}\,\mathrm{cm^3\,s^{-1}}$$

(Osterbrock & Ferland 2006). The inequality implies that it is possible that not all of the hydrogen is ionised.

By fitting a blackbody function to the 2.8 and 4.1 d spectra, we obtained the temperatures ($\approx$ 57 kK for 2.8 d and $\approx$ 21 kK for 4.1 d) and the radii (1.83 × 10$^{14}$ cm for 2.8 d and 4.24 × 10$^{14}$ cm for 4.1 d). Using these values and the H$\alpha$ line luminosity in Eqn. 1, we obtain K values: 5.5×10$^{15}$ and 3.4 ×10$^{15}$ g cm$^{-1}$. Here we have assumed that the CSM is attached to the progenitor and used a progenitor radius ($r_0$) of 1000 R$_\odot$.

Using an average value for K ($\approx$ 4.4×10$^{15}$ g cm$^{-1}$), gives a lower estimate of the mass-loss rate using

$$\dot{\mathrm{M}} = 4\pi \mathrm{K} v_{\mathrm{wind}},$$

which lies in the range 8.8 × 10$^{-4}$ to 8.8 × 10$^{-3}$ M$_\odot$ yr$^{-1}$, assuming $v_{\mathrm{wind}}$ to be 10 and 100 km s$^{-1}$, respectively.

### 5.4. Early spectra comparison with models

Dessart et al. (2017) characterised the spectroscopic signatures of RSG explosions embedded in an atmosphere/wind, extending to a maximum of 10 times the radius of the progenitor (R$_\star$) and having a mass $\leq$ 0.1 M$_\odot$, using a 1D non-local thermodynamic equilibrium (NLTE) radiative-transfer model, CMFGEN. They considered explosions of both compact RSGs (R$_\star$ = 501 R$_\odot$, 'r1') and extended RSGs (R$_\star$ = 1107 R$_\odot$, 'r2'), with different mass-loss rates ranging from 10$^{-6}$ to 10$^{-2}$ M$_\odot$ yr$^{-1}$. The wind terminal velocities in these models were fixed to 50 km s$^{-1}$.

We compared the early 2.8 and 4.1 d spectra of SN 2019nyk with the CMFGEN generated models. The optimal match for the 2.8 d spectrum was found with the 1.0 d r1w5r, 3.0 d r1w6, and 4.0 d r1w6 model spectra (see Fig. 15). Notably, the r1w6 model spectra are closer in epoch to the 2.8 d spectrum of SN 2019nyk.

The r1w5r model represents a compact RSG with a mass-loss rate of 5 × 10$^{-3}$ M$_\odot$ yr$^{-1}$ below 2 × 10$^{14}$ cm, transitioning to a lower mass-loss rate of 10$^{-6}$ M$_\odot$ yr$^{-1}$ beyond that radius. On the other hand, the r1w6 model corresponds to a compact RSG with a mass-loss rate of 10$^{-2}$ M$_\odot$ yr$^{-1}$, transitioning to a lower mass-loss rate of 10$^{-6}$ M$_\odot$ yr$^{-1}$ beyond 5 × 10$^{14}$ cm. The ejecta mass in both the models is 12.52 M$_\odot$, and the total masses of the material beyond R$_\star$ for r1w5r and r1w6 are 1.02 × 10$^{-2}$ and 3.04 × 10$^{-2}$ M$_\odot$, respectively.

All the three model spectra matches the features and continuum of the 2.8 d spectrum of SN 2019nyk quite well. However, the r1w5r model spectrum exhibits a higher He II flux, a lower C IV flux, and an additional feature at 5600 Å, absent in the observed spectrum. The 3.0 d r1w6 model spectrum displays an additional feature at 7100 Å and a higher flux for most of the features than those in the observed spectrum. On the other hand,

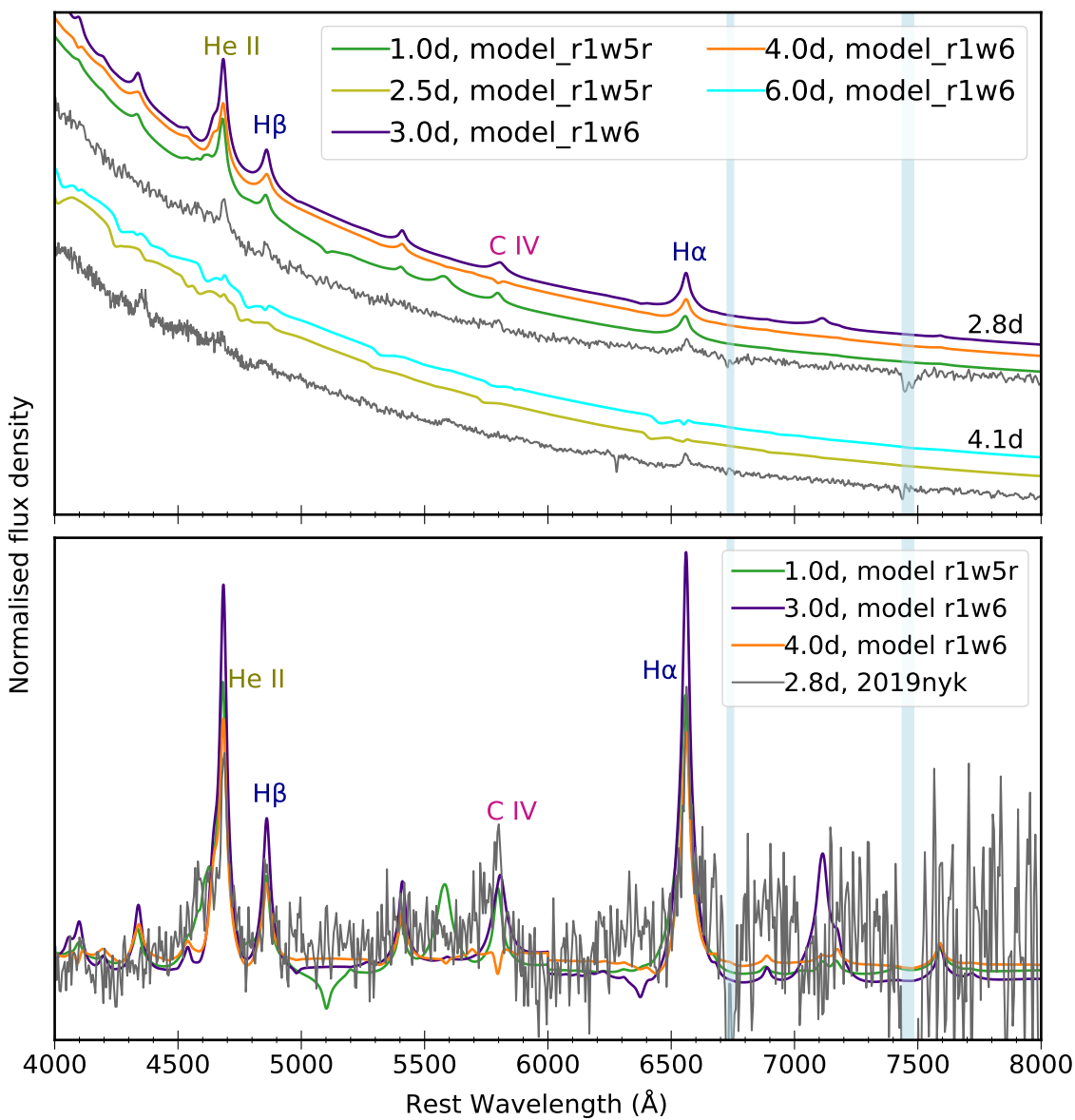

**Fig. 15.** Comparison of the 2.8 and 4.1 d median normalised spectra of SN 2019nyk with CMFGEN models from Dessart et al. (2017). The models are convolved with a Gaussian kernel to simulate the spectral resolution of the observations (R ≈ 390). The bottom panel shows the continuum normalised 2.8 d spectrum of SN 2019nyk and the model spectra.

the 4.0 d r1w6 model spectrum shows a higher He II flux and lower Hα flux, and no C IV feature.

For the 4.1 day spectrum of SN 2019nyk, although the continuum matches well with both the 2.5 d r1w5r and 6.0 d r1w6 model spectra, the features are not well-reproduced in either case.

We also carried out a comparative analysis of the *UBV(RI)ri* absolute magnitude and pseudo-bolometric LCs for SN 2019nyk with those of the models from Dessart et al. (2017) in Figure 16. We determined the pseudo-bolometric luminosity evolution using *UBVri* magnitudes for SN 2019nyk using SuperBol. To maintain consistency, the model *UBVRI* magnitudes from Dessart et al. (2017), simulated using HERACLES (González et al. 2007), were used to calculate model pseudo-bolometric LCs with SuperBol. The observed LCs closely resembled the shape of the LC models with an appended 'h', representing a compact RSG with an enhanced CSM mass close to the progenitor. The mass enhancement in these models near $R_\star$ implies stronger interaction, leading to higher bolometric luminosity compared to the models without mass enhancement. SN 2019nyk, however, displayed consistently higher luminosity across all bands and, consequently, in the pseudo-bolometric LC. Thus, despite comparison with model spectra suggests a less dense CSM around the progenitor, the higher luminosity of SN 2019nyk compared to the models necessitates the presence of a denser CSM.

Furthermore, we use the CMFGEN models as presented in Boian & Groh (2020) to investigate the properties of SN 2019nyk during its early phases. These models simulate events at epochs between ∼1-day to a few days after the explosion, assuming three different inner boundary radii: $R_{in}=8 \times 10^{13}$, $16 \times 10^{13}$ and $32 \times 10^{13}$ cm, corresponding to 1.0, 1.8 and 3.7 d post-explosion, respectively. These inner boundary radii indicate the extent to which the SN ejecta has swept up the CSM at a given time 't' after the explosion. In addition, the models explore a wide range of parameters: $1.9 \times 10^8 \leq L \leq 2.5 \times 10^{10}$ L$_\odot$, $10^{-3} \leq \dot{M} \leq 10^{-2}$ M$_\odot$ yr$^{-1}$, three values of chemical abundances (solar, He-rich, and CNO-processed), and fixed wind terminal velocities and ejecta expansion velocities of $v_\infty$ = 150 km s$^{-1}$ and $v_{ej}$=10$^4$ km s$^{-1}$, respectively.

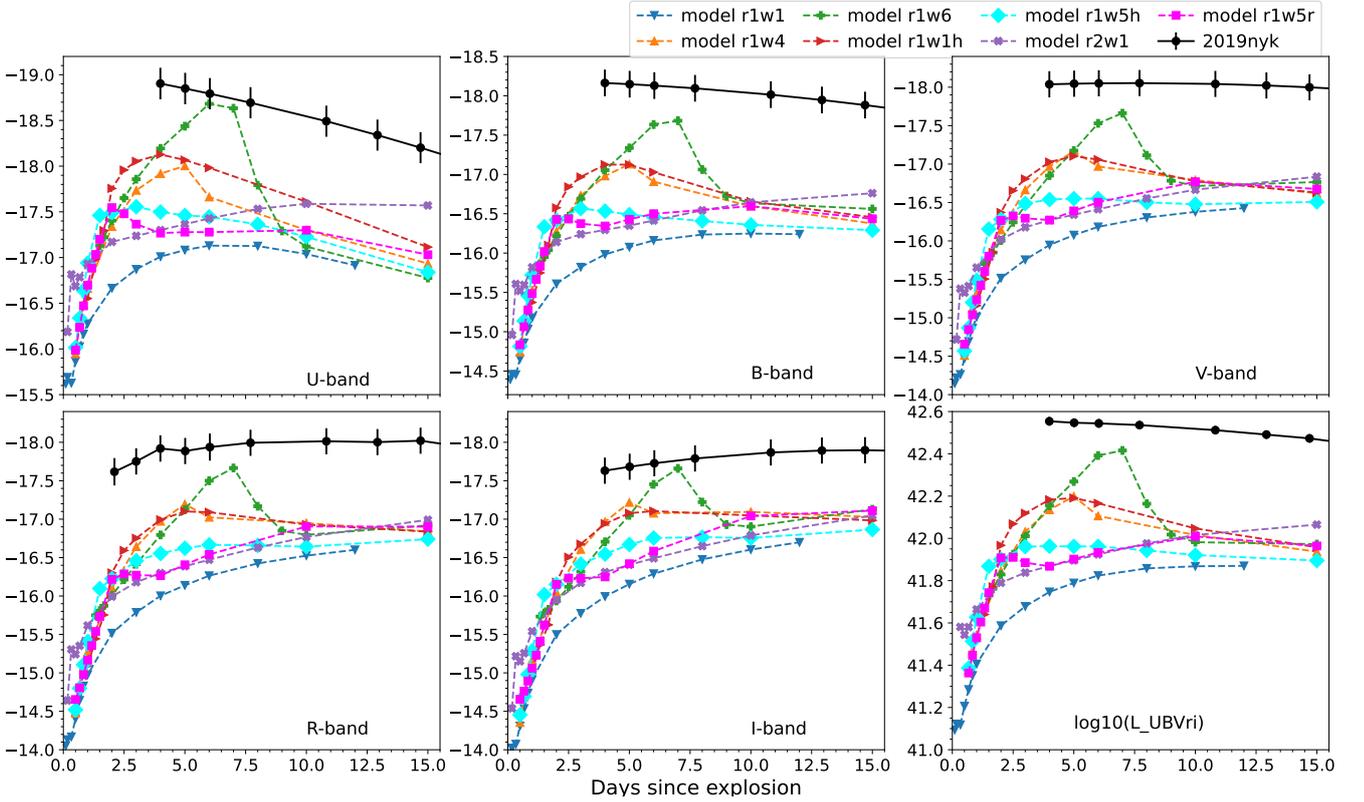

**Fig. 16.** Comparison of the multi-band absolute magnitude LCs and pseudo-bolometric luminosity evolution of SN 2019nyk to those of the models presented in Dessart et al. (2017). Here, Sloan-r and i band LCs of SN 2019nyk are compared to the Johnson RI band model LCs.

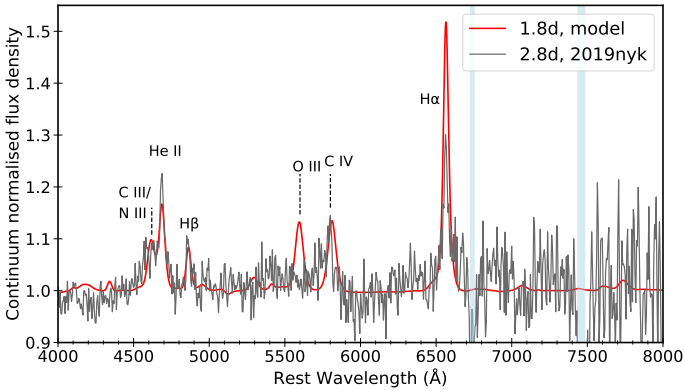

**Fig. 17.** Comparison of the 2.8 d continuum normalised spectrum of SN 2019nyk to the CMFGEN models from Boian & Groh (2020). The model is smoothed with a Gaussian kernel to emulate the spectral resolution of the observations (R ≈ 390).

In their study, Boian & Groh (2020) modelled 17 interacting SNe, concluding that in terms of surface abundances, the majority of these events are well-fitted by the assumption that the progenitors had CNO-processed surface abundances, as expected for evolved massive stars. A small number of events (15.7%) exhibited solar-like surface abundances, with another 26.3% displaying either solar-like or CNO-processed abundances.

For the 2.8 d spectrum of SN 2019nyk, we found the best match with 2.5 × 10$^{10}$ L$_\odot$ model (see Fig. 17), with $\dot{M}$ = 3×10$^{-3}$ M$_\odot$ yr$^{-1}$, an effective temperature (at op-tical depth = 10): $T_{eff}$ = 64.1 kK, and the mass fractions $X_H$ = 7.010×10$^{-1}$, $X_{He}$ = 2.800×10$^{-1}$, $X_C$ = 3.025×10$^{-3}$, $X_N$ = 1.182×10$^{-3}$, $X_O$ = 9.634×10$^{-4}$, and $X_{Fe}$ = 1.360×10$^{-3}$, corresponding to solar chemical abundance. The inner boundary is located at 1.6×10$^{14}$ cm.

Most of the features in the observed spectrum are present in the model. However, the best-fit model exhibits an O III feature, which is not apparent in the 2.9 d spectrum of SN 2019nyk. In contrast, a broad feature corresponding to O III wavelength is conspicuous in the 4.1 d spectrum of SN 2019nyk. Boian & Groh (2020) has suggested that the inability to simultaneously fit all the lines observed in the early spectra might indicate an aspherical geometry. The feature bluewards of He II could be attributed to either C III and/or N III. It is possible to rule out N III for the model corresponding to solar abundance composition as N emission is typically expected in CNO-processed material. Moreover, the CNO models exhibit a low abundance of C, thereby fails to reproduce the C IV feature effectively, which is observed in the spectrum of SN 2019nyk.

## 6. 1D radiation hydrodynamical modelling of LC

We use the open-source 1D radiation hydrodynamics code, Supernova Explosion Code (SNEC, Morozova et al. 2015) for multi-band LC modelling in order to infer the progenitor parameters and explosion properties of SN 2019nyk. SNEC is a local thermodynamic equilibrium (LTE) code that invokes grey opacities without spectral calculations. SNEC takes progenitor model, explosion energy, $^{56}$Ni mass and $^{56}$Ni mixing as input and simulate a range of outputs, including multi-band LCs, bolometric LC, photospheric velocity, and temperature evolution. To investigate the influence of CSM on the LCs, various densi-ties and radial extents of CSM can be added on top of the progenitor model. For the progenitor, we adopt models from Sukhbold et al. (2016), which are computed using the stellar evolution code KEPLER (Weaver et al. 1978). These models represent progenitors evolved until core collapse from a range of non-rotating, solar composition ZAMS stars spanning masses from 9 to 120 M$_\odot$. We generated a grid of SNEC models encompassing ZAMS masses between 11 and 16 M$_\odot$, explosion energy within 1.0 - 1.5 foe, while maintaining a constant $^{56}$Ni mass of 0.046 M$_\odot$. The $^{56}$Ni mixing parameter, representing the mass-coordinate until which $^{56}$Ni is mixed outwards, is fixed to 3 M$_\odot$. Since ejecta-CSM interaction boosts the early luminosity of the LC, we compared the epochs from 30 d until $t_{PT}$ of the observed LCs with the $Vri$ band model LCs without CSM. We did not include the bluer band LCs, that is, $UBg$ bands, as these LCs are more susceptible to be affected by line blanketing at later phases. Based on chi-square minimisation, we identify the best-fit solution corresponds to a 13.0 M$_\odot$ ZAMS star, with a pre-SN mass of 11.57 M$_\odot$, a H-envelope mass less than 5.64 M$_\odot$ and a pre-SN stellar radius of 700 R$_\odot$.

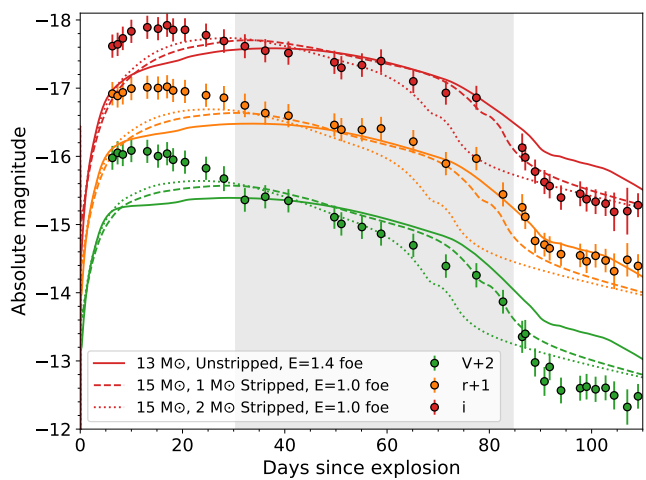

**Fig. 18.** Unstripped, 1 M$_\odot$, and 2 M$_\odot$ stripped multi-band LC (without CSM) SNEC models are shown with solid, dashed, and dotted lines, respectively. The shaded region denotes the range of epochs within which the fits are performed.

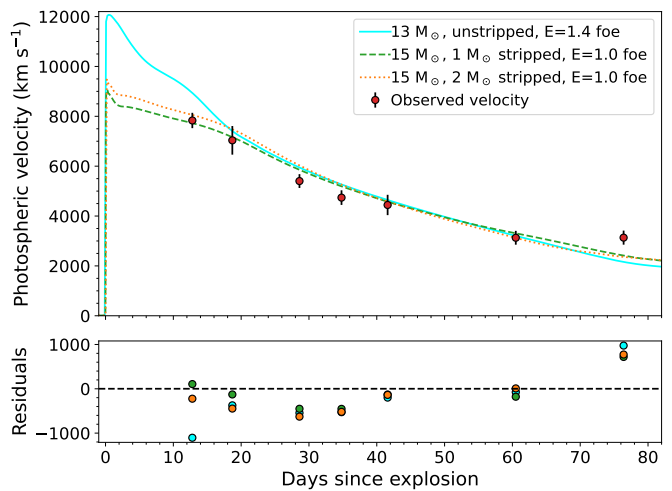

**Fig. 19.** Photospheric velocity evolution corresponding to the three models shown in Fig. 18 along with the observed photospheric velocity.

Furthermore, we used the stripped-mSGB series progenitor models from Morozova et al. (2015), where the ZAMS stars are evolved until core collapse using MESA (Paxton et al. 2019; Jermyn et al. 2023, and references therein). The methodology behind these models is as follows: a ZAMS star with a mass of 15 M$_\odot$ is evolved initially until the middle of the sub-giant branch phase (mSGB) when its effective temperature ($T_{eff}$) reaches $10^4$ K. At this juncture, a portion of the star's mass is instantaneously removed, after which the evolution continues until the onset of core collapse. These stripped-mSGB models encompass a range of mass removal scenarios, leading to varied outcomes. Specifically, models with less than 6 M$_\odot$ of mass removed result in RSG progenitors, as determined by the $T_{eff}$ criterion established by Georgy (2012). On the other hand, stripping in the range of 6 - 7 M$_\odot$ yields progenitors classified as yellow supergiants. In Fig. 18, we present the unstripped model, corresponding to the explosion of the progenitor models from Sukhbold et al. (2016), alongside the 1 M$_\odot$ and 2 M$_\odot$ stripped models, demonstrating their multi-band LCs that best match the epochs from 30 d to 83 d ($t_{PT}$ for SN 2019nyk). As expected, the degree of stripping influences the slope and the duration of the plateau phase in the LC. More heavily stripped progenitors exhibit steeper and shorter plateaus and the effect of stripping on the LC slope is more prominent in the bluer bands. Among the three models, the 1 M$_\odot$ stripped progenitor aligns most closely with the $Vri$ band LCs of SN 2019nyk. This particular progenitor model corresponds to a pre-SN mass of 11.27 M$_\odot$, an envelope mass of less than 6.18 M$_\odot$, and a pre-SN stellar radius of 1031 R$_\odot$. As far as the velocity evolution is concerned, the model photospheric velocity after 30 days is similar in all three models, as shown in Fig. 19. Given these considerations, we proceeded with the 1 M$_\odot$ stripped progenitor model for further exploration of the effects of CSM in our analysis.

For modelling the early LC (< 30d), we attached CSM at the surface of the progenitor, assuming a spherical wind density profile for the CSM around the progenitor. The density profile has the same form as described in Section 5.3. Consequently, when CSM is integrated into the model, the mass loading parameter, $K_{csm}$, and the CSM extent, $R_{csm}$, become two additional free parameters. We generated a grid of models spanning an explosion energy ($E_{exp}$) range of 0.8 to 1.3 foe in increments of 0.1 foe. The mass loading parameter ($K_{csm}$) was varied between 2 to $10 \times 10^{17}$ g cm$^{-1}$ in intervals of $1 \times 10^{17}$ g cm$^{-1}$, while the CSM radius ($R_{csm}$) was explored from 1400 to 4000 R$_\odot$ in steps of 100 R$_\odot$. In addition, given the uncertainty in the explosion time, we shift the observed LCs by ±2 d in steps of 1 d and treat this offset as a free parameter. As earlier, the $^{56}$Ni mass and $^{56}$Ni mixing parameters are fixed to 0.046 M$_\odot$ and 3 M$_\odot$, respectively. We note that the selected range of values for the mass-loading parameter inherently results in a density discontinuity at the interface between the progenitor model and the connected CSM, and we have applied artificial smoothing to this transition.

We employed chi-square minimisation technique to obtain the model that simultaneously best fit the observed $Vri$ band LCs from explosion until $t_{PT}$ and the observed photospheric velocity evolution. The best-fit model corresponds to an explosion energy of 1.0 foe, a CSM radial extent of 2800 R$_\odot$ and a mass-loading parameter of $6 \times 10^{17}$ g cm$^{-1}$. In Fig. 20, the observed multi-band LCs are displayed alongside the best-fitting model. The best-fit model successfully reproduces the early LC (< 30d) in the $B$ and $g$ bands. However, the model LCs deviates from the observed data in the $U$-band and beyond 30 days in the $B$ and $g$ bands, possibly due to the unaccounted effects of line blanketing in these bands in the modelling.

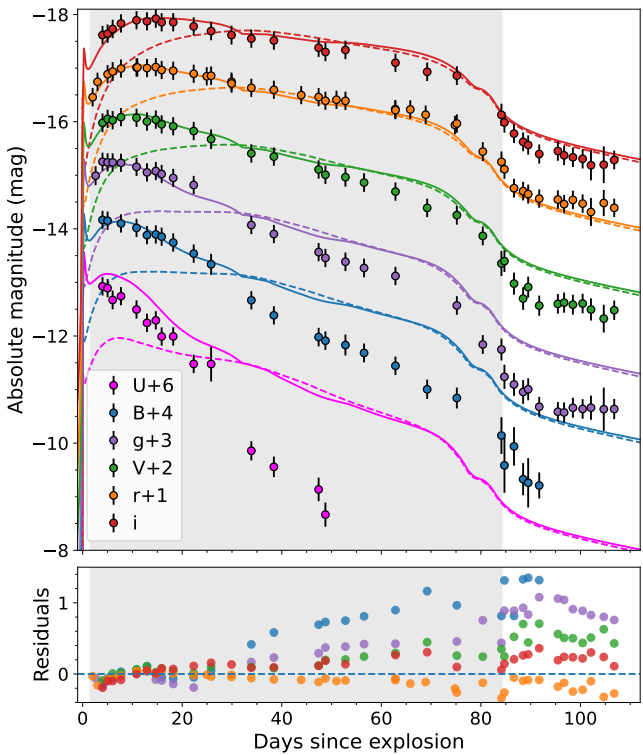

**Fig. 20.** Best-fitted multi-band LC SNEC models with CSM are shown with solid lines. The fits are performed with respect to *Vri* bands luminosity until $t_{PT}$. The model without CSM is shown with dashed lines.

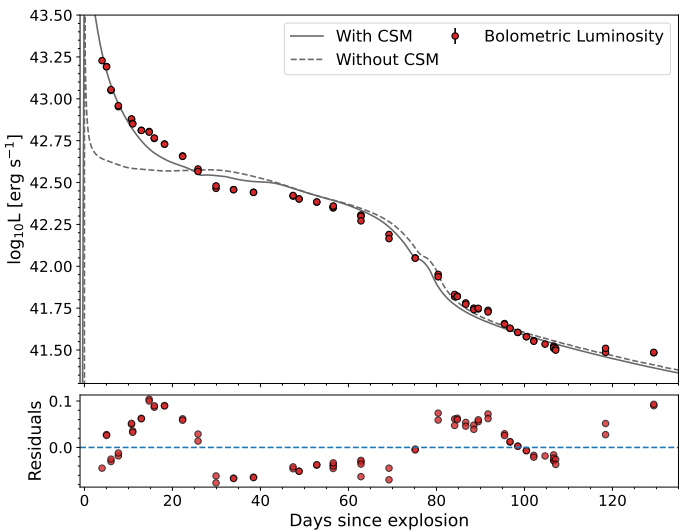

**Fig. 21.** Best-fitted model bolometric luminosity, corresponding to a simultaneous fit to the observed bolometric light curve and the photospheric velocity, is shown with a solid line. The dashed line denotes the best-fitted model without CSM.

Considering the impact of line blanketing on the bluer bands, it is imperative to determine the best model for the bolometric LC. To achieve this, we used the blackbody-corrected luminosity estimated with SuperBol (Nicholl 2018) using *BgVri* photometry as the observed bolometric luminosity of SN 2019nyk. Employing chi-square minimisation, we obtained the best-fit model that simultaneously fit the observed bolometric LC and the velocity, which is characterised by an explosion energy of 1.1 foe, a CSM radial extent of 2800 $R_\odot$, and a mass-loading parameter of $3 \times 10^{17}$ g cm$^{-1}$. Fig. 21 presents the best-fit model bolometric luminosity evolution alongside the observed bolometric luminosity.

In Fig. 22, we present the best-fit model photospheric velocities achieved when fitting multi-band LCs and velocities simultaneously, along with the case where bolometric light curves and velocities were fitted simultaneously. These are compared with the observed photospheric velocities. The observed photospheric velocity at each epoch corresponds to the velocity of the element with the lowest velocity at that time. The residuals clearly indicate that the models associated with the simultaneous best fit to bolometric light curves and velocities provide a closer match to the observed velocities.

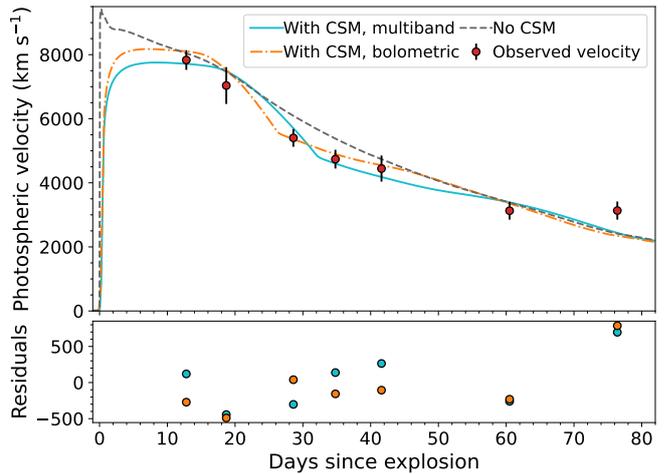

**Fig. 22.** Best-fit model photospheric velocities achieved when fitting multi-band light curves and velocities simultaneously, along with the case where bolometric light curves and velocities were fitted simultaneously are over-plotted on the observed photospheric velocity.

To derive parameter uncertainty estimates, a set of 1000 absolute magnitude *Vri* band LCs was generated, using distances drawn randomly from a normal distribution, with mean as 94.7 Mpc and 7.4 Mpc as the standard deviation. For each of these LCs, the best-fit parameters were estimated, leading to the generation of 1000 data points for explosion energy, CSM radius, and mass-loading parameter and time offset. The posterior distribution of the parameters and their correlations are shown in Fig. 23. The parameter values correspond to the median of the posterior distribution, which represents the best-fitting solution.

To establish uncertainty limits, we calculated the 95th percentile of the marginalised probability density function and then subtracted the 50th percentile to determine the upper error. Similarly, the lower error was determined by subtracting the 5th percentile from the 50th percentile. This choice is made in favour of mean and standard deviation values, as not all posterior distributions conform to a Gaussian distribution. Therefore, opting for the median and percentiles provides us with more robust insights into the distribution. The parameter values and their uncertainties are, therefore, $E_{exp} = 1.1 \pm 0.1$ foe, $R_{csm} = 2900^{+800}_{-300}$ $R_\odot$, and $K_{csm} = 4^{+3}_{-2} \times 10^{17}$ g cm$^{-1}$. These values are in agreement with those obtained via fitting the multi-band and the bolometric LCs within errors.

The CSM mass can be written as

$$M_{csm} = \int_{R_{in}}^{R_{csm}} 4\pi \rho_{csm}(r) r^2 dr \qquad (2)$$

where $R_{in}$ is the inner CSM radius and $R_{csm}$ is the outer CSM radius.

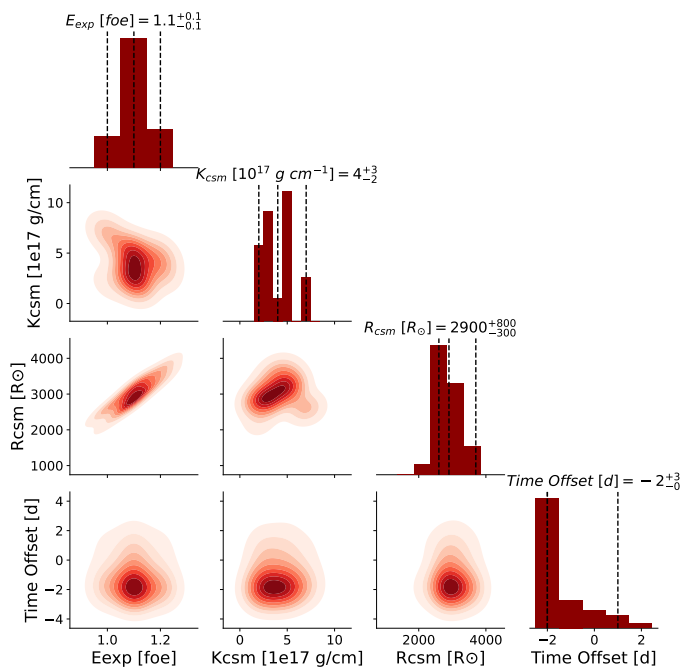

**Fig. 23.** Corner plot displaying 1D and 2D posterior distributions for parameter uncertainty estimation. The diagonal sub-plots show 1D histograms of the posterior distribution of the explosion energy ($E_{exp}$), the mass loading parameter ($K_{csm}$), the CSM radial extent ($R_{csm}$) and the time offset. The 5th, 50th, and 95th percentiles for each of the parameters are shown with vertical dashed lines. The off-axis sub-plots show 9 contour levels for 2D joint probability densities at various confidence intervals, ranging from 10% to 90%.

Substituting $\rho_{csm}(r) = K_{csm} r^{-2}$ and replacing $R_{in}$ with the progenitor radius as the CSM here is assumed to be attached to the progenitor, and integrating, we get

$$M_{csm} = 4\pi K_{csm}(R_{prog} - R_{csm}).$$

Here, $R_{prog}$ represents the progenitor radius. From this expression, we calculated the total CSM mass to be $0.16^{+0.14}_{-0.09}\,M_\odot$, using $R_{prog} = 1031\,R_\odot$, $R_{csm} = 2900^{+800}_{-300}\,R_\odot$, and $K_{csm} = 4^{+3}_{-2} \times 10^{17}\,\mathrm{g\,cm^{-1}}$. While it is possible to estimate the mass-loss rate from the CSM mass, given a reasonable wind speed, it is important to note that SNEC modelling does not allow for a precise determination of the mass-loss rate or the duration of the outflow. This is because these models primarily depend on the density profile of the wind and are independent of the wind velocity in the pre-explosion models. Consequently, a wind speed corresponding to steady-state RSG winds of approximately 10 km s$^{-1}$ would imply a lower mass-loss rate for the progenitor. On the other hand, higher wind velocities, around 100 km s$^{-1}$, would correspond to increased mass-loss rates on the order of a few $M_\odot$ yr$^{-1}$ shortly before the explosion.

## 7. Discussion and conclusions

In this paper, we have conducted a comprehensive analysis of the rapidly declining Type II supernova, SN 2019nyk. The LC properties of SN 2019nyk is similar to those of other fast-declining Type II SNe, such as SNe 2013by and 2014G. The luminosity of SN 2019nyk rises to a peak absolute magnitude of $-18.09 \pm 0.17$ mag in the *V* band, and exhibits a decline rate of $2.84 \pm 0.03$ mag (100 d)$^{-1}$ in the recombination phase. The recombination phase lasts until ~75 days followed by a 1.4 mag drop in *V* band magnitude. In the radioactive tail phase, a short 'plateau tail' phase in the *g* and *V* bands has been observed, reminiscent of the prototypical Type II SN 1999em.

The early spectral observations of SN 2019nyk exhibit the presence of high-ionisation narrow emission lines, including He II and C IV, as well as narrow H Balmer emission lines. We compare these features with those of other SNe that show early interaction and find a resemblance between these features and those observed in SNe 2014G and 2023ixf. These features are believed to originate from the unshocked part of a slowly moving CSM. While the narrow emission lines disappear after 4.1 d, the 20.2 d spectrum clearly shows a boxy H$\alpha$ emission profile with very weak absorption, which is a signature of ejecta-CSM interaction.

We employed both radiation hydrodynamic modelling for the LC (SNEC) and comparisons with established literature LC and spectral models to constrain the nature of the CSM, specifically obtaining the mass-loading parameter, $K_{csm}$, which is directly proportional to the density of the CSM. We emphasise that SNEC is an 1D LTE gray radiation hydrodynamics code, and it differs significantly from CMFGEN, which is a NLTE radiative transfer code and has been employed to produce spectral models found in literature. The H$\alpha$ line luminosity during the early phases indicates a lower limit for $K_{csm}$ of $4.4 \times 10^{15}$ g cm$^{-1}$. Further comparisons with CMFGEN models from Dessart et al. (2017) have constrained the parameter $K_{csm}$ to $5 \times 10^{15}$ g cm$^{-1}$. When considering CMFGEN models by Boian & Groh (2020), the derived value of $K_{csm}$ from the best-matching model with the 2.8 d observed spectrum of SN 2019nyk is $1 \times 10^{15}$ g cm$^{-1}$. Additionally, the latter model suggests a solar composition surface abundance for the progenitor. We note that the wind velocities used in the CMFGEN models in the two studies are different (50 km s$^{-1}$ in Dessart et al. 2017 and 150 km s$^{-1}$ in Boian & Groh 2020). Finally, our LC modelling yields a value for $K_{csm}$ of $4^{+3}_{-2} \times 10^{17}$ g cm$^{-1}$, which is about two orders of magnitude higher than those derived from spectral model comparisons. This implies that LC modelling suggests a denser CSM compared to estimates based on H$\alpha$ line luminosity and spectral comparisons.

The CSM parameters obtained from SNEC modelling align with the compact wind scenario (e.g. Haynie & Piro 2021), suggesting that the shock breakout is expected within the CSM rather than at the stellar surface. However, a potential limitation arises when considering a CSM with $K_{CSM}$ of $4 \times 10^{17}$ g cm$^{-1}$ extending to a $R_{CSM}$ of 2900 $R_\odot$ (~ $2 \times 10^{14}$ cm). By adopting the photospheric radius obtained from fitting a blackbody function to the 2.8 d spectrum as the shock breakout radius ($R_s = 1.8 \times 10^{14}$ cm $\approx$ 2600 $R_\odot$) and integrating the optical depth expression from $R_s$ to $R_{CSM}$, we obtain a large electron scattering optical depth of $\approx$ 80, assuming an electron scattering opacity $\kappa$ of 0.34 cm$^2$ g$^{-1}$. Thus, the mean free path of recombination photons responsible for forming the 'IIn-like' features is much smaller than $R_{CSM}$. Consequently, the CSM characteristics inferred from SNEC alone cannot account for the 'IIn-like' features, and a less dense CSM would be necessary to generate them. On the other hand, while existing spectral models from Dessart et al. (2017) employing low CSM density could replicate the narrow features in the early spectra, the corresponding LC models fail to match the early-time luminosity of SN 2019nyk. The higher luminosity of SN 2019nyk supports the presence of a denser CSM than those in the models. Moreover, we note that the explosion energy for SN 2019nyk estimated from LC modelling (1.1 foe) is lower than that used in Dessart et al. (2017) models (1.35 foe). For models with lower explosion energy, the mismatch would be more pronounced, indicating that an even denser CSM would be required in the models to reproduce the observed LCs.

This discrepancy may be reconciled by introducing low-density material above the high density CSM derived from SNEC modelling. In the context of SN 2019nyk, this implies an overall CSM with a density profile deviating from $r^{-2}$, encompassing SNEC-inferred high-density material ($K_{CSM} \sim 1\times10^{17}$ g cm$^{-1}$) with a sharp transition to low-density material ($K_{CSM} \sim 1\times10^{15}$ g cm$^{-1}$) at $R_{CSM} > 2\times10^{14}$ cm (e.g. see Dessart & Jacobson-Galán 2023). Another explanation, as proposed by Morozova et al. (2017), is that the CSM is non-spherical. In this case, the SNEC-inferred CSM component represents regions where the shock can pass into the CSM, while the low-density material above and below the SNEC-inferred CSM facilitates the formation of narrow lines. Alternatively, instead of a steady state wind, the CSM may resemble an accelerated wind (Moriya et al. 2018). In such a scenario, the CSM could be more extended, and $K_{csm}$ could be lower.

While CSM density can be estimated from LC modelling, the determination of the mass-loss rate is not definitive. Mass-loss rate estimates rely on the wind velocity, which can be challenging to obtain due to the limited resolution of early spectra. Assuming a standard RSG wind velocity of 10 km s$^{-1}$, and using the value of $K_{csm}$ from LC modelling, we infer a mass-loss rate of approximately 0.08 $M_\odot$ yr$^{-1}$. The associated duration of the wind is given by

$$t_{wind} = \frac{R_{csm} - R_{prog}}{v_{wind}},$$

which implies that the mass loss event would have commenced around 4.1 years prior to the explosion. Alternatively, if we assume a super-wind velocity of 100 km s$^{-1}$ (e.g. Smith et al. 2023), the mass-loss rate would be around 0.8 $M_\odot$ yr$^{-1}$, indicating that the CSM formation from mass loss began less than 4.9 months before the explosion. Moreover, we can also estimate duration of mass loss from the epoch of disappearance of narrow features from the spectra of SN 2019nyk, which is 4.1 d after explosion. Assuming the ejecta velocity at 4.1 d is similar to the H$\beta$ FWHM velocity at that epoch, which is 2800 km s$^{-1}$, the mass loss duration would range between 3.1 years to 3.8 months prior to explosion. These values correspond to wind velocities of 10 km s$^{-1}$ and 100 km s$^{-1}$, respectively, and align with the findings derived from LC modelling. Finally, the progenitor and explosion parameters estimated from the LC modelling are the following: pre-SN mass = 11.27 $M_\odot$, an envelope mass of less than 6.18 $M_\odot$, a pre-SN stellar radius of 1031 $R_\odot$, explosion energy $E_{exp}$ = 1.1 foe and radial extent of CSM, $R_{csm}$ = 2900 $R_\odot$.

The early detection of numerous transient events, driven by high cadence all-sky surveys, has enabled the identification of many young transients displaying IIn-like features. However, it is crucial for future studies to obtain high-resolution early spectra of these young transients (e.g. Smith et al. 2023). This is essential for accurately constraining the wind velocity, a critical parameter for estimating the rate of mass loss and determining the temporal profile of the mass-loss phenomenon. Improved measurements of the time-frame of the mass-loss event can establish connections between these events and distinct stages of stellar burning, thereby revealing the underlying origin of the CSM (Smith & Arnett 2014; Woosley & Heger 2015).

## Acknowledgments

We express our gratitude to the anonymous referee for providing us with valuable suggestions and scientific insights that enhanced the quality of the paper. RD acknowledges funds by ANID grant FONDECYT Postdoctorado N° 3220449. G.P. acknowledges support by ANID grant FONDECYT regular N° 1201793. This work makes use of data from the Las Cumbres Observatory global network of telescopes. The LCO group is supported by NSF grants AST-1911151 and AST-1911225. We acknowledge Wiezmann Interactive Supernova data REPository http://wiserep.weizmann.ac.il (WISeREP, Yaron & Gal-Yam 2012). This research has made use of the NASA/IPAC Extragalactic Database (NED) which is operated by the Jet Propulsion Laboratory, California Institute of Technology, under contract with the National Aeronautics and Space Administration. We acknowledge the usage of the HyperLeda database (http://leda.univ-lyon1.fr).